\documentclass[a4paper,11pt]{article}
\pdfoutput=1
\usepackage{jheppub}
\usepackage[utf8]{inputenc}
\usepackage[T1]{fontenc}
\usepackage{amsmath,amssymb,amsfonts,bm}
\usepackage{graphicx}
\usepackage{xcolor}
\usepackage{tikz}
\usetikzlibrary{arrows.meta,decorations.markings,decorations.pathmorphing,positioning,calc}
\usepackage[compat=1.1.0]{tikz-feynman}
\numberwithin{equation}{section}

\newcommand{\dd}{\mathrm d}
\newcommand{\Tr}{\mathrm{Tr}}

\newcommand{\HT}{H_T}

\newcommand{\calN}{\mathcal N}

\newcommand{\bluerev}[1]{{\color{blue}#1}}

\tikzset{
  dirac/.style={line width=0.95pt,postaction={decorate},decoration={markings,mark=at position 0.50 with {\arrow{Latex[length=2.1mm,width=1.4mm]}}}},
  diracplain/.style={line width=0.95pt},
  scalar/.style={dashed,thick},
  boson/.style={dashed,thick},
  hline/.style={dashed,thick},
  philine/.style={dashed,thick},
  vertex/.style={circle,fill=black,inner sep=1.45pt},
  op/.style={rectangle,draw,thick,inner sep=2pt},
  flexural/.style={line width=0.9pt,decorate,decoration={snake,amplitude=0.55mm,segment length=2.7mm}},
  membranevertex/.style={circle,fill=black,inner sep=1.7pt},
  strainvertex/.style={rectangle,draw=black,fill=white,line width=0.85pt,minimum width=0.75cm,minimum height=0.48cm,inner sep=1.5pt},
  diagramtitle/.style={font=\normalsize},
  panellabel/.style={font=\normalsize},
  every picture/.style={baseline=-0.5ex,line cap=round,line join=round}
}

\title{\boldmath Critical Ripples and Dirac Fermions in Crystalline Membranes}
\author[a]{Sebasti\'an Bahamondes}
\author[b]{Rodrigo Soto-Garrido}
\author[b]{Enrique Mu\~noz}
\author[c]{Vladimir Juri\v{c}i\'c}

\affiliation[a]{Instituut-Lorentz (LION), Universiteit Leiden,
Einsteinweg 55, 2333 CC Leiden, The Netherlands}
\affiliation[b]{Facultad de F\'isica, Pontificia Universidad Cat\'olica de Chile,
Vicu\~na Mackenna 4860, Santiago, Chile}
\affiliation[c]{Departamento de F\'isica, Universidad T\'ecnica Federico Santa Mar\'ia,
Casilla 110, Valpara\'iso, Chile}

\emailAdd{bahamondes@lorentz.leidenuniv.nl}
\emailAdd{rodrigo.sotog@gmail.com}
\emailAdd{ejmunozt@uc.cl}
\emailAdd{juricic@gmail.com}

\abstract{%
Crystalline membranes hosting Dirac fermions, with graphene as the
paradigmatic example, combine two low-energy sectors with sharply
different dynamics: nonrelativistic flexural phonons and
relativistic-like Dirac quasiparticles. We develop the low-energy field
theory of this coupled system at charge neutrality and determine how
this dynamical mismatch controls the coupling between the two sectors.
In the long-wavelength flat phase, rotational symmetry ties the renormalization of the leading local
scalar strain--density coupling to the scale-dependent bending
rigidity, causing its dimensionless strength
to decrease logarithmically.
At the same time, flexural modes become parametrically slower than
Dirac fermions, so the resulting fermionic feedback vanishes as a
power law. The flat phase is therefore stable against this
perturbation. The physics changes when
elastic interactions or electronic softening destabilize the membrane
at a finite wavelength, selecting a ripple pattern formed by modes at
$\pm\mathbf{Q}$. For an isolated pair of ordering wavevectors, provided that
commensurability-induced phase pinning is irrelevant, the transition
is governed by the bosonic Wilson--Fisher fixed point, while the
Dirac fermions remain spectators. A genuinely hybrid
electronic--structural critical point arises instead when symmetry
permits a mass-type Dirac bilinear to share the ripple's momentum and other
quantum numbers, including horizontal-reflection parity. The transition
is then described by the chiral-XY Gross--Neveu--Yukawa universality
class. Using the known one-loop critical exponents, we characterize this
transition, determine the induced secondary elastic distortion, and
show that the fermionic and bosonic velocities lock in the isotropic
continuum limit.
}


\begin{document}
\maketitle
\flushbottom

\section{Introduction}
\label{sec:intro}

Dirac materials provide condensed-matter realizations of relativistic quantum field theories, with the Fermi velocity playing the role of an effective speed of light \cite{CastroNeto2009,Wehling2014,Armitage2018}.  They provide a versatile setting for exploring quantum criticality,
including dynamical mass generation, emergent symmetries, and anomalous
scaling of observables.   In particular, interacting Dirac systems provide prominent examples of emergent relativistic scaling~\cite{HerbutJuricicRoy2009,HerbutJuricicVafek2009}, supported by both numerical simulations~\cite{AssaadHerbut2013PRX,OtsukaYunokiSorella2016PRX} and experimental observations~\cite{Ma2025RelativisticMott}.
Furthermore, long-range electromagnetic interactions can
drive the Dirac velocity toward the propagation velocity of the
gauge field~\cite{Gonzalez1994,Isobe-Nagaosa-PRB2012}, while short-range interactions at fermionic quantum
critical points, including Gross--Neveu--Yukawa (GNY) transitions, can
lock the fermionic and order-parameter velocities to an emergent  common
terminal value~\cite{RoyHerbut2016,RoyKennettYangJuricic2018,ReiserJuricic2024}.

Here we consider a qualitatively different situation, in which relativistic-like quasiparticles are coupled to nonrelativistic collective modes.  Our system is a crystalline membrane hosting massless Dirac fermions, as realized in graphene and related two-dimensional Dirac materials \cite{MarianiOppen2008,KatsnelsonFasolino2013}.  After the in-plane displacement field is integrated out, the flexural sector is governed by a quadratic dispersion and a nonlocal elastic interaction.  The membrane therefore has dynamical exponent $z=2$, whereas for the Dirac quasiparticles  $z=1$.  The coupled theory consequently lacks a common microscopic spacetime scaling, and its infrared structure must be analyzed while retaining this mismatch explicitly.

The flat phase of a crystalline membrane is stabilized by anharmonic elasticity, which produces an infrared enhancement of the bending rigidity \cite{NelsonPeliti_JPhysique_1987,Aronovitz_PRL_1988,LeDoussalRadzihovsky1992,Kownacki_PRE_2009,Coquand_PRE_2020,MetayerTeber_JSTAT_2025,MauriKatsnelson2022,NelsonPiranWeinberg2004,LeDoussalRadzihovsky2018,KatsnelsonFasolino2013}.  Coupling the membrane to  Dirac fermions introduces a competing tendency to softening, since the electronic density response generates an attractive nonlocal strain interaction.  The central question is therefore whether this electronic softening affects the long-wavelength stability of the flat membrane or becomes irrelevant in the infrared.

At charge neutrality we focus on the local scalar interaction between the conserved Dirac density $\rho_\psi$ and the isotropic flexural-strain operator $\HT$, defined in eq.~\eqref{eq:Sg}. The operator $\HT$ is conjugate to isotropic tension. In the absence of membrane--fermion coupling, the rotational Ward identity fixes the linearized scaling dimension of $\HT$ through the field and bending-rigidity renormalizations~\cite{NelsonPeliti_JPhysique_1987,Aronovitz_PRL_1988,Kownacki_PRE_2009,MauriKatsnelson2022}; see appendices~\ref{app:strain} and~\ref{app:quantum}.  Fermion corrections to the mixed vertex are not protected by this Ward identity, but power counting shows that they are governed by the irrelevant feedback coupling $\alpha_e$.  The resulting renormalization group (RG) flow causes the dimensionless strength of this leading local scalar interaction to decrease logarithmically at long wavelengths, while the fermion-induced nonlocal quartic strain interaction is irrelevant.  The  polarization does not generate overdamped flexural dynamics, and flexural fluctuations do not modify the leading infrared scaling of the Dirac kinetic term.  The long-wavelength flat phase is therefore stable against the local perturbation in eq.~\eqref{eq:Sg}.

This asymptotic weak-coupling result does not exclude a nonuniversal instability at stronger bare coupling or at an intermediate momentum scale.  It is therefore compatible with earlier proposals for electron-induced rippling and curvature condensation \cite{GonzalezPerfetto2009,SanJoseGonzalezGuinea2011,Gonzalez2014,Guinea_PRB_2014}.  Those scenarios concern finite-momentum softening that occurs before the asymptotic flat-membrane regime is reached, whereas our results establish the universal infrared behavior of the local scalar flexural interaction within that regime. The distinction between asymptotic irrelevance and a possible finite-$Q$ instability underlies the classification developed below.

Finite-momentum rippling defines a separate set of effective theories once microscopic elasticity, lattice anisotropy, substrate effects or electronic feedback produces isolated minima at $\pm{\bf Q}$.  When no low-energy Dirac bilinear carries both the momentum and the microscopic quantum numbers of the ripple field, the leading fermionic perturbation is the coupling between the fermion  and boson order-parameter density $\lambda|\Phi_Q|^2\rho_\psi$.  Charge conservation fixes $\Delta_\rho=D$, while the Wilson--Fisher correlation-length exponent gives $\Delta_{|\Phi|^2}=D+1-1/\nu$.  The perturbing operator therefore has scaling dimension
\begin{equation}
\Delta_{|\Phi_Q|^2\rho_\psi}
=
2D+1-\frac{1}{\nu},
\end{equation}
yielding the scaling dimension of the coupling  
$y_\lambda=1/\nu-D$.   
For the two-component ripple order parameter describing the modes at
$\pm{\bf Q}$, the three-dimensional XY value of $\nu$ gives
$y_\lambda<0$ in $D=2$, and the perturbation is irrelevant.
The transition is consequently governed by the bosonic Wilson--Fisher fixed point, with the Dirac fermions remaining spectators at criticality, although the same coupling shifts the Dirac-point energy in the ordered phase.

A qualitatively different critical theory arises when the finite-momentum structural field transforms as a mass-type Dirac bilinear under the full microscopic symmetry group, including horizontal reflection.  In this case a local Yukawa coupling is allowed, and the critical mode becomes a hybrid electronic--structural order parameter.  In the simplest two-component realization the resulting theory belongs to the chiral-XY GNY universality class.  The explicit graphene valley--sublattice construction presented in section~\ref{sec:gny} realizes the required pair of Kekul\'e mass matrices.  For a mirror-symmetric free-standing sheet, however, a pure out-of-plane ripple is mirror odd whereas the conventional Kekul\'e bond mass is mirror even, so a linear coupling between them is forbidden.  The graphene realization therefore requires broken horizontal reflection or a critical distortion containing a symmetry-compatible bond component.  Under these conditions, the known chiral-XY critical exponents characterize the transition, an electronic instability induces a secondary structural distortion through Landau mixing, and the fermionic and bosonic velocities flow toward a common value within the isotropic continuum theory.

Our central result is a field-theoretic classification of
membrane--Dirac criticality based on dynamical scale separation and
microscopic symmetry: the flat phase asymptotically decouples,
whereas finite-momentum ordering realizes either Wilson--Fisher
criticality with spectator Dirac fermions or a hybrid
electronic--structural GNY transition.

The paper is organized as follows.  In section~\ref{sec:lowenergy} we introduce the effective membrane--fermion theory and the scaling conventions used in the flat and finite-momentum problems.  Section~\ref{sec:flat} analyzes the flat membrane coupled to  Dirac fermions.  In section~\ref{sec:finiteQ} we formulate the mechanical finite-$Q$ transition, while section~\ref{sec:gny} treats the symmetry-locked GNY route.  We present our conclusions in section~\ref{sec:conclusions}.  Technical details are collected in appendices~\ref{app:feynman}--\ref{app:brazovskii}.

\section{Low-energy theory and scaling regimes}
\label{sec:lowenergy}

We start by defining the effective fields and the scaling conventions used throughout the paper.  The long-wavelength flat membrane, the Dirac fermions and finite-momentum order-parameter fields involve different upper critical dimensions and different dynamical scalings.  We keep these regimes separate from the outset, since the flat-phase deformation potential and the finite-$Q$ Yukawa theory are controlled by distinct power-counting rules.

We parametrize the infrared RG flow by
\begin{equation}
 \ell=\ln\frac{\Lambda}{\mu_{\rm RG}},
 \qquad
 \dot g\equiv\frac{\dd g}{\dd\ell},
\end{equation}
where $\Lambda$ is the ultraviolet cutoff and $\mu_{\rm RG}$ is the
running renormalization scale, which decreases toward the infrared.
We use the integration conventions
\begin{equation}
 \int_{\omega,{\bf k}}
 \equiv
 \int\frac{\dd\omega}{2\pi}
 \int\frac{\dd^D{\bf k}}{(2\pi)^D},
 \qquad
 \int_{\tau,{\bf x}}
 \equiv
 \int\dd\tau\,\dd^D{\bf x}.
\end{equation}

\subsection{Quantum membrane and Dirac fermions}

We consider a $D$-dimensional crystalline membrane embedded in $D+d_c$ dimensions.  Its out-of-plane fluctuations are described by $d_c$ height fields $h_a(\tau,{\bf x})$.  After the in-plane displacement field is integrated out \cite{NelsonPeliti_JPhysique_1987},
the flexural effective action is most transparently written in momentum space as

\begin{align}
 S_h={1\over2}\int_{\omega,{\bf k}}h_a(-\omega,-{\bf k})
 \left(\rho\omega^2+\kappa_b k^4\right)h_a(\omega,{\bf k})
 +{1\over4}\int_{\tau,{\bf q}}\mathcal U_{ij}({\bf q})
 R_{ij,kl}({\bf q})\mathcal U_{kl}(-{\bf q}),
\label{eq:Sh}
\end{align}
and 
\begin{equation}
 \mathcal U_{ij}({\bf q})={1\over2}\int_{\bf k}
 k_i(q-k)_j h_a(\tau,{\bf k})h_a(\tau,{\bf q}-{\bf k}).
\label{eq:Uijq}
\end{equation}

The resulting elastic kernel is nonlocal and contains the momentum-dependent transverse projector

\begin{equation}
 P^T_{ij}({\bf q})=\delta_{ij}-{q_iq_j\over q^2}.
\label{eq:PTproj}
\end{equation}
We denote by $M_{ij,kl}({\bf q})$ and $N_{ij,kl}({\bf q})$ the orthogonal dilatational and traceless projectors constructed from $P^T({\bf q})$, 
\begin{eqnarray}
M^{jklr}(\mathbf{q})&=&\frac{1}{2}\left(P_{jl}^{\,T}(\mathbf{q})P_{kr}^{\,T}(\mathbf{q})+P_{jr}^{\,T}(\mathbf{q})P_{kl}^{\,T}(\mathbf{q})\right)-N^{jklr}(\mathbf{q}),\\
N^{jklr}(\mathbf{q})&=&\frac{1}{D-1}P_{jk}^{\,T}(\mathbf{q})P_{lr}^{\,T}(\mathbf{q}),
\end{eqnarray}
and write
\begin{equation}
 R_{ij,kl}({\bf q})=\mu M_{ij,kl}({\bf q})+Y N_{ij,kl}({\bf q}).
\label{eq:Rdecomp}
\end{equation}

The compact notation suppresses the momentum dependence of the
projectors; all contractions in appendix~\ref{app:strain} are
understood in the corresponding transverse projected basis. Our
analysis focuses on the local trace coupling $H_T\rho_\psi$.
Integrating out the in-plane displacement also generates a
momentum-projected mixed strain vertex and an induced
density--density interaction. These nonlocal terms require a separate
composite-operator analysis and are not included here.

We also include massless Dirac fermions with action
\begin{equation}
 S_\psi=\int_{\tau,{\bf x}}\bar\psi_\alpha
 \left(\gamma^0\partial_\tau+v_F\gamma^i\partial_i\right)\psi_\alpha,
\label{eq:Spsi}
\end{equation}
with $\alpha=1,\ldots,N_f$, $\Tr\,\mathbf{1}=N_D$, where $N_D$ is the dimension of the Dirac representation, and $\bar\psi=\psi^\dagger\gamma^0$. Here, $\{\gamma^\mu,\gamma^\nu\}=2\delta^{\mu\nu}$.  The local scalar flexural interaction studied here couples the conserved Dirac density $\rho_\psi$ to the metric trace $\HT$,

\begin{equation}
 S_g=g\int_{\tau,{\bf x}}\HT\rho_\psi,
 \qquad
 \HT={1\over2}\delta_{ij}\partial_i h_a\partial_j h_a,
 \qquad
 \rho_\psi=\bar\psi\gamma^0\psi .
\label{eq:Sg}
\end{equation}
For graphene-like systems, the number of two-component cones entering the density response is
\begin{equation}
 \calN={N_fN_D\over2}.
\label{eq:Ncone}
\end{equation}

For $N_f=2$ and $N_D=4$, one has $\calN=4$.

\subsection{Three scaling problems}
\label{sec:threeRG}

Three scaling regimes enter the analysis.  Throughout, $D$ denotes the spatial (internal membrane) dimension; the three upper critical dimensions $D=4$ (classical), $D=2$ (quantum) and $D=3$ (relativistic) are all spatial, the relativistic one referring to a spacetime dimension $D+1=4$.  The first is the classical static membrane, used below as a benchmark for the tensor algebra of the flat phase.  It has a propagator $G_h({\bf k})\sim k^{-4}$ and upper critical spatial dimension $D=4$.  The second is the zero-temperature quantum membrane.  Its propagator is the $z=2$ kernel in eq.~\eqref{eq:Sh}; the frequency integration lowers the upper critical spatial dimension to $D=2$.  The third is the isolated finite-$Q$ order-parameter theory.  It is relativistic at long wavelengths and has upper critical spatial dimension $D=3$.

We use the dimensional regulators
\begin{equation}
 \epsilon_m=4-D\, \text{(classical membrane)},
 \,\, {\rm and} \,\,
 \epsilon=3-D\, \text{(relativistic finite-}Q\text{ sectors)} .
\label{eq:epsdefs}
\end{equation}
In sections~\ref{sec:finiteQ}--\ref{sec:gny} and appendices~\ref{app:WF},~\ref{app:gny}, the symbol $\epsilon$ means $3-D$, while $\epsilon_m=4-D$ is reserved for the classical membrane benchmark.  The flat-phase coupling joins sectors with $z_h=2$ and $z_\psi=1$, and is treated by weighted power counting around the flexural scaling $\omega\sim k^2$.  We keep the coefficient $\kappa_b$ of $k^4$ explicit and introduce

\begin{equation}
 \bar g={g\over\kappa_b},\qquad
 s(k)={\sqrt{\kappa_b/\rho}\,k\over v_F},\qquad
 \alpha_e(k)={\calN\over16}\bar g^{\,2}s
 ={\calN\over16}{g^2k\over v_F\rho^{1/2}\kappa_b^{3/2}}.
\label{eq:mixedcouplings}
\end{equation}

The variable $s$ quantifies the hierarchy between the energy scales associated with  the flexural elastic and Dirac fermion modes, while $\alpha_e$ is the dimensionless strength of the nonlocal quartic strain interaction generated by the  fermion polarization bubble.  
Both have negative tree-level scaling dimensions at the flexural fixed
point and therefore flow to zero in the infrared, whereas the quantum
Young modulus is only marginally irrelevant. The finite-$Q$ couplings are instead analyzed within the relativistic-like theories of sections~\ref{sec:finiteQ} and~\ref{sec:gny}.

\section{Flat membrane coupled to Dirac fermions at charge neutrality}
\label{sec:flat}

We now analyze the flat phase.  Throughout this section we study the local flexural-metric coupling in eq.~\eqref{eq:Sg}.  As explained in section~\ref{sec:lowenergy}, this is the local trace-strain vertex whose infrared scaling is controlled by the tension Ward identity (see appendix~\ref{app:strain}). The classical membrane theory first shows how rotational symmetry constrains the scaling of the isotropic-tension operator. The zero-temperature problem must then be treated separately because the flexural and Dirac sectors have different dynamical scalings.

\subsection{Scaling of the tension operator in the classical regime}

The one-loop beta functions of the classical membrane are standard \cite{NelsonPeliti_JPhysique_1987,Aronovitz_PRL_1988,LeDoussalRadzihovsky1992,Kownacki_PRE_2009,Coquand_PRE_2020,MetayerTeber_JSTAT_2025,BowickTravesset2001,Gazit2009,MauriKatsnelson2020}, and take the form 
\begin{align}
 \dot\mu&=\epsilon_m\mu-{\mu\over 6(4\pi)^2}\left[(d_c+20)\mu+10Y\right],
\label{eq:betamu}\\
 \dot Y&=\epsilon_mY-{Y\over 6(4\pi)^2}\left[20\mu+\left(10+{5d_c\over2}\right)Y\right],
\label{eq:betaY}\\
 \eta_h&={5(2\mu+Y)\over6(4\pi)^2}+O(\mu^2,Y^2).
\label{eq:etah}
\end{align}
The infrared stable fixed point is
\begin{equation}
 \mu_*=(4\pi)^2{6\epsilon_m\over d_c+24},\qquad
 Y_*=(4\pi)^2{12\epsilon_m\over5(d_c+24)},
\label{eq:memFP}
\end{equation}
with
\begin{equation}
 \eta_h^*={12\epsilon_m\over d_c+24}.
\label{eq:etastar}
\end{equation}
For the physical codimension $d_c=1$ and $D=2$, the classical benchmark gives $\eta_h^*=24/25$.

An isotropic tension $\sigma$ couples to the scalar strain $\HT$ through $S_\sigma=\sigma\int\HT$.  The rotational Ward identity of the tensionless membrane implies that the divergent part of the flexural self-energy begins at order $p^4$ \cite{NelsonPeliti_JPhysique_1987,Aronovitz_PRL_1988,MauriKatsnelson2022,RoldanFasolinoZakharchenkoKatsnelson2011,BurmistrovStress2018,BurmistrovDifferential2018}.  The coefficient of the $\sigma p^2$ insertion is therefore fixed by the field and bending-rigidity renormalizations.  Appendix~\ref{app:strain} gives the Ward-identity argument, while appendix~\ref{app:quantum} gives the corresponding zero-temperature one-loop check. The scaling dimension of the tension   is
\begin{equation}
 y_\sigma=2-\eta_h. 
\label{eq:ysigma}
\end{equation}
Combining this static tension exponent with the dimension of the conserved density gives  for the scaling dimension of the coupling,
\begin{equation}
 y_g^{\rm cl}=y_\sigma-D=2-D-\eta_h.
\label{eq:ygclassical}
\end{equation}
At the classical membrane fixed point this becomes
\begin{equation}
 y_g^{\rm cl}=2-D-{12(4-D)\over d_c+24}.
\label{eq:ygclassicalFP}
\end{equation}
This result defines a static one-loop benchmark for the scaling
dimension of the local tension coupling. For $D=2$ and $d_c=1$, the
direct one-loop extrapolation gives $y_g^{\rm cl}=-24/25$, indicating
irrelevance within the static regime. Determining the corresponding
scaling dimension in the full dynamical membrane--Dirac theory requires
an analysis of the complete momentum-projected deformation-potential
vertex.

\begin{figure}[t]
\centering
\begin{tikzpicture}[x=1.1cm,y=1.1cm,line cap=round,line join=round]

\begin{scope}[shift={(0,3.35)}]
  \draw[flexural] (-1.55,0)--(-0.12,0);
  \draw[flexural] (0.12,0)--(1.55,0);
  \node[membranevertex] at (0,0) {};
  \draw[flexural]
    (-0.02,0.08) .. controls (-0.68,0.92) and (0.68,0.92) .. (0.02,0.08);
  \node[panellabel] at (0,-0.72) {(a)};
\end{scope}

\begin{scope}[shift={(6.0,3.35)}]
  \node[membranevertex] (VL) at (-0.72,0) {};
  \node[membranevertex] (VR) at (0.72,0) {};
  \draw[flexural] (VL)--(-1.85,0.72);
  \draw[flexural] (VL)--(-1.85,-0.72);
  \draw[flexural] (VR)--(1.85,0.72);
  \draw[flexural] (VR)--(1.85,-0.72);
  \draw[flexural]
    (VL) .. controls (-0.28,0.72) and (0.28,0.72) .. (VR);
  \draw[flexural]
    (VR) .. controls (0.28,-0.72) and (-0.28,-0.72) .. (VL);
  \node[panellabel] at (0,-1.02) {(b)};
\end{scope}

\begin{scope}[shift={(0,0)}]
  \node[strainvertex] (H) at (0,0) {$\HT$};
  \draw[flexural] (-1.65,0)--(H.west);
  \draw[flexural] (H.east)--(1.65,0);
  \draw[flexural]
    (H.north west) .. controls (-0.70,1.00) and (0.70,1.00) .. (H.north east);
  \node[panellabel] at (0,-0.78) {(c)};
\end{scope}

\begin{scope}[shift={(6.0,0)}]
  \node[strainvertex] (HL) at (-1.35,0) {$\HT$};
  \node[strainvertex] (HR) at (1.35,0) {$\HT$};
  \draw[flexural] (-2.45,0.55)--(HL.north west);
  \draw[flexural] (-2.45,-0.55)--(HL.south west);
  \draw[flexural] (HR.north east)--(2.45,0.55);
  \draw[flexural] (HR.south east)--(2.45,-0.55);
  \draw[dirac]
    (HL.east) .. controls (-0.62,0.78) and (0.62,0.78) .. (HR.west);
  \draw[dirac]
    (HR.west) .. controls (0.62,-0.78) and (-0.62,-0.78) .. (HL.east);
  \node[panellabel] at (0,-1.02) {(d)};
\end{scope}

\end{tikzpicture}
\caption{Flat-phase one-loop Feynman diagrams.  Wiggly lines denote flexural propagators and solid directed lines denote Dirac propagators.  Panels (a) and (b) generate the pure-membrane renormalizations in eqs.~\eqref{eq:betamu}--\eqref{eq:etah}.  Panel (c) is the isotropic-tension insertion entering eqs.~\eqref{eq:ysigma} and~\eqref{eq:betagbar}; its Ward-identity derivation is given in appendices~\ref{app:strain} and~\ref{app:quantum}.  Panel (d) generates the nonlocal strain interaction in eq.~\eqref{eq:nonlocalstrain}, with the polarization given in eq.~\eqref{eq:PiDyn}.}
\label{fig:flatgraphs}
\end{figure}
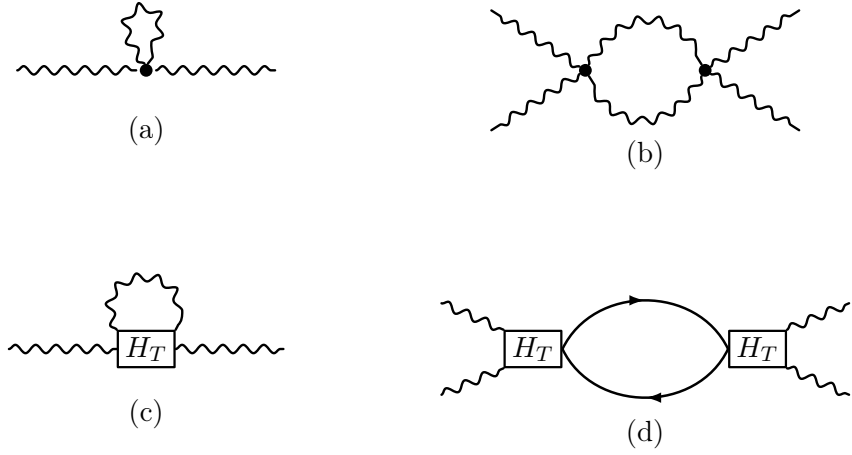

The one-loop structures underlying the membrane renormalization, the tension insertion, and the fermion-induced strain interaction are summarized in figure~\ref{fig:flatgraphs}.

\subsection{\texorpdfstring{Zero-temperature scaling and asymptotic decoupling}
{Zero-temperature scaling and asymptotic decoupling}}
\label{sec:Tzero}

At zero temperature, the flexural dispersion is quadratic,
$\omega_h(k)=\sqrt{\kappa_b/\rho}\,k^2$, and the corresponding dynamical
exponent is $z_h=2$. Power counting therefore identifies $D=2$ as the upper
critical spatial dimension of the quantum membrane. The elastic interaction
may be parametrized by the dimensionless quantum Young coupling
\begin{equation}
 y=\frac{Y}{\rho^{1/2}\kappa_b^{3/2}},
\label{eq:yquantum}
\end{equation}
Here $\kappa_b$ is the bending-rigidity coefficient introduced in eq.~\eqref{eq:Sh}. The RG flow of the pure quantum membrane was derived in
refs.~\cite{KatsLebedev2014,Amorim2014,MauriKatsnelson2022}. In the
normalization adopted in appendix~\ref{app:quantum}, it reads
\begin{align}
 \frac{\dd y}{\dd\ell}
 &=-a y^2+O(y^3),
\label{eq:betayq}\\
 \frac{\dd\ln\kappa_b}{\dd\ell}
 &=b y+O(y^2),
\label{eq:betakappaq}
\end{align}
with
\begin{equation}
 a=\frac{3(d_c+6)}{128\pi},
 \qquad
 b=\frac{3}{32\pi}.
\end{equation}
The corresponding  solutions are
\begin{equation}
 y(\ell)
 =\frac{y_0}{1+a y_0\ell},
\label{eq:qmemsolution}
\end{equation}
and 
\begin{equation}
 \kappa_b(\ell)
 =\kappa_{b0}
   \left[1+a y_0\ell\right]^\theta,
\label{eq:kappaSolution}
\end{equation}
where $\theta=b/{a}={4}/(d_c+6)$. 
Thus the elastic interaction is marginally irrelevant, while the bending
rigidity increases logarithmically toward the infrared.

Rotational invariance constrains the renormalization of the local
isotropic-tension insertion. Since isotropic tension is the source conjugate
to $\HT$, the rotational Ward identity implies that pure-membrane ultraviolet
divergences in the flexural self-energy begin at order $p^4$ rather than
$p^2$~\cite{MauriKatsnelson2022}. The Ward-identity argument is given in
appendix~\ref{app:strain}, where the zero-momentum $\HT$ insertion is obtained
by differentiating the flexural two-point function with respect to tension;
see eqs.~\eqref{eq:selfenergyfactor} and~\eqref{eq:insertionDerivative}.
Appendix~\ref{app:quantum} verifies this result explicitly at one loop:
eqs.~\eqref{eq:quantumSelfSchematic}--\eqref{eq:insertionUVfinite} show that
the $p^2$ insertion has no ultraviolet pole, yielding
$Z_H^{\rm(vtx)}=1$ in eq.~\eqref{eq:ZHquantum}.

Charge conservation implies that the conserved density has no
independent renormalization. Equivalently, the Ward
identity fixes its scaling dimension to
$\Delta_{\rho_\psi}=D$~\cite{PeskinSchroeder1995}; in the present normalization this is
expressed as $Z_\rho=1$. After the canonical
rescaling in eqs.~\eqref{eq:canonicalRescaling}
and~\eqref{eq:canonicalMixedParams}, the dimensionless local coupling is
\begin{equation}
 \bar g=\frac{g}{\kappa_b}.
\end{equation}
Combining $Z_H^{\rm(vtx)}=Z_\rho=1$ with the bending-rigidity flow in
eq.~\eqref{eq:qRGapp} therefore gives
\begin{equation}
 \frac{\dd\ln\bar g}{\dd\ell}
 =-b y+O(y^2,\alpha_e).
\label{eq:betagbar}
\end{equation}
This Ward-identity result applies to the zero-momentum tension insertion.
Throughout this section, $\bar g$ denotes the coefficient of the corresponding
leading scalar projection of the local vertex. At finite momentum transfer,
derivative operators may mix with $\HT$ and require a separate
composite-operator analysis. To leading logarithmic order,
\begin{equation}
 \bar g(\ell)
 =\bar g_0
 \left[1+a y_0\ell\right]^{-\theta}.
\label{eq:gbarSolution}
\end{equation}

The flexural and Dirac sectors have different dynamical scalings. Their
frequency ratio at the running momentum $k(\ell)=\Lambda e^{-\ell}$ is
\begin{equation}
 s(\ell)
 =\frac{\sqrt{\kappa_b(\ell)/\rho}\,k(\ell)}
        {v_F(\ell)}.
\label{eq:sratio}
\end{equation}
The same quantity appears as the expansion parameter of the fast Dirac
propagator in eq.~\eqref{eq:fastFermionExpansion}. Moreover, the fermion
self-energy derived in eq.~\eqref{eq:fermionSelfScaling} contains no
$p\ln p$ term, so the Dirac velocity has no leading infrared logarithmic
renormalization and satisfies
$\dd\ln v_F/\dd\ell=O(\alpha_e)$. Using the bending-rigidity flow then gives
\begin{align}
 \frac{\dd\ln s}{\dd\ell}
 &=-1+\frac{b}{2}y+O(y^2,\alpha_e),
\nonumber\\
 s(\ell)
 &=s_0 e^{-\ell}
   \left[1+a y_0\ell\right]^{\theta/2}.
\label{eq:sflow}
\end{align}
Hence $s(\ell)\to0$: at a fixed long-wavelength momentum, the flexural modes
become parametrically slower than the Dirac fermions,
$\omega_h\ll\omega_\psi$.

This scale separation permits a controlled treatment of the mixed problem.
For external flexural kinematics $\Omega\sim q^2$, the fermionic determinant
can be expanded in $s\sim\Omega/(v_Fq)$, as shown explicitly in
eq.~\eqref{eq:fastFermionExpansion}, producing effectively instantaneous
strain interactions. Their normalization follows from the Dirac
density--density polarization function at neutrality (zero chemical potential) derived in
appendix~\ref{app:polarization} and given in
eq.~\eqref{eq:PiAppResult}. The leading fermionic feedback is
therefore measured by
\begin{equation}
 \alpha_e
 =\frac{\calN}{16}\bar g^{\,2}s
 =\frac{\calN}{16}
   \frac{g^2 k}
        {v_F\rho^{1/2}\kappa_b^{3/2}}.
\label{eq:alphae}
\end{equation}
Using eqs.~\eqref{eq:betagbar} and~\eqref{eq:sflow}, one obtains
\begin{align}
 \frac{\dd\ln\alpha_e}{\dd\ell}
 &=-1-\frac{3b}{2}y+O(y^2,\alpha_e),
\label{eq:alphaeflow}\\
 \alpha_e(\ell)
 &=\alpha_{e0}e^{-\ell}
   \left[1+a y_0\ell\right]^{-3\theta/2}.
\label{eq:alphaesolution}
\end{align}
These asymptotics are also summarized in
eq.~\eqref{eq:mixedAsymptoticsApp}. The electronic feedback is therefore
power-law irrelevant, with an additional logarithmic suppression from
membrane fluctuations.

To determine whether the fermions renormalize the local coupling $g$ in
eq.~\eqref{eq:Sg}, we project the corresponding 1PI vertex onto the
tree-level tensor structure,
\begin{equation}
 \Gamma^{\rm 1PI}_{hh\bar\psi\psi}
 (\mathbf p_1,\mathbf p_2;P)
 =
 (\mathbf p_1\!\cdot\!\mathbf p_2)\gamma^0
 \left[g_R+\delta g_R(P)\right]
 +\Gamma_{\rm der},
\label{eq:mixedVertexDecomposition}
\end{equation}
where $\Gamma_{\rm der}$ contains operators with additional external
derivatives. The corresponding one-loop calculation is given in
eqs.~\eqref{eq:bareMixedVertex} and~\eqref{eq:mixedVertexOneLoop}.
At $P=0$, the order-$g^2$ contribution vanishes because its integrand is odd
under $(\omega,\mathbf k)\mapsto(-\omega,-\mathbf k)$. The first surviving
term is proportional to $P_\mu$ and belongs to $\Gamma_{\rm der}$ rather
than to the renormalization of the original local coupling.

The first possible correction to the original local tensor structure
contains an additional pair of normalized density--strain vertices and
is therefore of order $\bar g^{\,2}$. As  shown in appendix~\ref{app:quantum}, its local projection is
suppressed by at least one power of $s=\omega_h/(v_Fk)$. The fermionic
contribution therefore satisfies the infrared bound
\begin{equation}
 \gamma_g^{(\psi)}
 =O\!\left(\bar g^{\,2}s\right)
 =O(\alpha_e),
\label{eq:mixedVertexPowerMain}
\end{equation}
with possible additional powers of $s$. Since $\alpha_e\to0$ in the
infrared, the fermionic vertex correction vanishes asymptotically, and
the leading infrared flow of the local coupling remains determined by
the pure-membrane Ward identity.

The converse influence of the membrane on the Dirac fermions is also
subleading. Appendix~\ref{app:quantum} derives the full trace-strain
correlator in eqs.~\eqref{eq:strainCorrelatorConvolution}
and~\eqref{eq:strainCorrelatorScaling}, together with the logarithmic bound
in eq.~\eqref{eq:strainCorrelatorBound}. This result implies that the fermion
self-energy takes the form 
\begin{equation}
 \Sigma_\psi^{\rm IR}(P)
 =O\left(
 \frac{g^2}{\rho^{1/2}\kappa_b^{3/2}}
 p^2\ln\frac{\Lambda}{p}
 \right),
\end{equation}
as shown in eq.~\eqref{eq:fermionSelfScaling}. In particular, no
$p\ln p$ term is generated, so the Dirac kinetic term retains its canonical
infrared scaling.

Equations~\eqref{eq:gbarSolution}, \eqref{eq:alphaesolution},
\eqref{eq:mixedVertexPowerMain}, and~\eqref{eq:fermionSelfScaling}
therefore establish the asymptotic decoupling of the flexural and Dirac
sectors at charge neutrality. We next evaluate the Dirac density response
to determine the normalization and dynamical structure of the induced
nonlocal strain interaction.

\subsection{Dirac polarization and undamped flexural dynamics}
\label{sec:nodamping}

The density bubble at charge neutrality is \cite{WunschStauberSolsGuinea2006,HwangDasSarma2007}
\begin{equation}
 \Pi_{00}(i\Omega,{\bf q})
 =-{\calN\over16}{q^2\over\sqrt{v_F^2q^2+\Omega^2}}.
\label{eq:PiDyn}
\end{equation}
In the static limit this becomes
\begin{equation}
 \Pi_{00}(0,{\bf q})=-{\calN\over16v_F}|q|.
\label{eq:PiStatic}
\end{equation}

For graphene, $\calN=4$, and one recovers $\Pi_{00}(0,q)=-|q|/(4v_F)$.  Because $\HT$ is quadratic in the height field, integrating out the Dirac fermions at order $g^2$ generates a nonlocal quartic strain interaction,

\begin{equation}
 \Delta S_{\rm nl}={g^2\over2}\int_{\Omega,{\bf q}}
 \HT(-\Omega,-{\bf q})\Pi_{00}(i\Omega,{\bf q})
 \HT(\Omega,{\bf q}).
\label{eq:nonlocalstrain}
\end{equation}
whose static kernel is proportional to $-\calN|q|/(16v_F)$, so the induced term softens the quartic strain
interaction.  At charge neutrality, the leading dynamical correction starts at order $\Omega^2/|q|$, reflecting the vanishing density of states and yielding undamped flexural dynamics.  Expanding in the flexural kinematic regime, $\Omega\sim q^2\ll v_Fq$, gives
\begin{equation}
 \Pi_{00}(i\Omega,{\bf q})=-{\calN\over16v_F}|q|
 +{\calN\over32v_F^3}{\Omega^2\over |q|}+\cdots .
\label{eq:PiExpand}
\end{equation}

The frequency-dependent correction scales as $q^3$ under $z=2$ and is subleading to the static nonlocal strain kernel.  Its strength relative to the quantum flexural action is precisely the running coupling $\alpha_e$ in eq.~\eqref{eq:alphae}, which vanishes according to eq.~\eqref{eq:alphaesolution}.  Dirac fermions at charge neutrality therefore leave the strain sector undamped and preserve the asymptotic quantum-membrane flow at weak coupling.


%
%

%
%

\section{Finite-momentum order parameter}
\label{sec:finiteQ}

We next consider a rippling instability at a nonzero wavevector, which
requires a microscopic mechanism to produce isolated minima. In the finite-momentum sector, we specialize to a single soft
flexural polarization, $d_c=1$, and assume that compression, substrate
coupling, competing elastic interactions, lattice anisotropy, or electronic
feedback produces a quadratic kernel $\mathcal K_h(\mathbf k)$ with minima at
the lattice-selected momenta $\pm\mathbf Q$,

\begin{equation}
 \mathcal K_h({\bf Q}+{\bf q})=r+c_\parallel^2q_\parallel^2
 +c_\perp^2q_\perp^2+\cdots .
\label{eq:finiteQkernel}
\end{equation}

In the absence of a symmetry-allowed linear coupling to gapless fermions, the inverse propagator near an isolated minimum is analytic in $\Omega^2$ and in deviations from $\mathbf Q$, yielding the relativistic finite-$Q$ theory used below.

For this isolated pair the height field can be written as
\begin{equation}
 h(\tau,{\bf x})=\Phi_Q(\tau,{\bf x})e^{i{\bf Q}\cdot{\bf x}}
 +\Phi_Q^*(\tau,{\bf x})e^{-i{\bf Q}\cdot{\bf x}},
\label{eq:hPhi}
\end{equation}
where the complex amplitude $\Phi_Q(\tau,\mathbf x)$ varies slowly
on the scale $Q^{-1}$.  The resulting field  theory is

\begin{equation}
 S_\Phi=\int_{\tau,{\bf x}}\left[|\partial_\tau\Phi_Q|^2+c_\parallel^2|\partial_\parallel\Phi_Q|^2+c_\perp^2|\nabla_\perp\Phi_Q|^2
 +r|\Phi_Q|^2+{\widetilde\lambda_4\over2}|\Phi_Q|^4\right].
\label{eq:SphiComplex}
\end{equation}
For comparison with the $O(N_b)$ convention used below, we write
$\Phi_Q=(\Phi_1+i\Phi_2)/\sqrt{2}$. The dimensionful real-field quartic coefficient $\lambda_4$ in $\lambda_4(\Phi_a^2)^2/4!$ is then related to eq.~\eqref{eq:SphiComplex} by
$\lambda_4=3\widetilde\lambda_4$. At zero temperature, the relativistic theory has
$d=D+1=4-\epsilon$ spacetime dimensions. We define the dimensionless
coupling
\begin{equation}
 u={\mu_{\rm RG}^{-\epsilon}\lambda_4\over8\pi^2}.
\label{eq:WFcouplingDef}
\end{equation}

The point-group star of ${\bf Q}$ is the set of ordering wavevectors related to ${\bf Q}$ by the lattice point group.  When this star consists only of $\pm{\bf Q}$, the reality of the height field makes the two amplitudes complex conjugates, so a single complex field $\Phi_Q$ suffices.  A lattice translation by ${\bf a}$ acts on this field  as $\Phi_Q\to e^{i{\bf Q}\cdot{\bf a}}\Phi_Q$.  For an incommensurate wavevector these phase rotations become an emergent $U(1)$ symmetry of the long-wavelength theory, and the phase of $\Phi_Q$ is the phason coordinate.  For a commensurate wavevector satisfying $n{\bf Q}={\bf G}$, with ${\bf G}$ a reciprocal lattice vector, translation symmetry permits the Umklapp term \cite{McMillan1976,Bak1982}

\begin{equation}
 \delta S_{\rm lock}=w_n\int_{\tau,{\bf x}}
 \left(\Phi_Q^n+\Phi_Q^{*n}\right),
\label{eq:lockin}
\end{equation}
which reduces the emergent phase symmetry to $\mathbb Z_n$ and generally pins the phason.  If ${\bf Q}=-{\bf Q}$ modulo a reciprocal lattice vector, the order parameter can be chosen real; if the point group generates several inequivalent pairs, the order parameter field  contains one complex component for each pair.  For the incommensurate single-pair case considered below, the ordered phase supports a gapless phason.  Denoting by $Q=|{\bf Q}|$ and by $u_s$ the local displacement of the modulation along ${\bf Q}$, we write

\begin{equation}
 \Phi_Q={1\over2}(A+\delta A)e^{-iQu_s},
\label{eq:phasesig}
\end{equation}
where $\delta A$ is the massive amplitude fluctuation and $u_s$ is the phason displacement.

  A nearly continuous shell of minimizing wavevectors instead leads to the Brazovskii problem discussed in appendix~\ref{app:brazovskii} \cite{Brazovskii1975,SwiftHohenberg1977}. The mechanical route applies when elasticity or lattice effects select ${\bf Q}$ and the ripple transforms in a representation distinct from those of the low-energy Dirac bilinears.  Symmetry then makes the coupling between the conserved charge density and the ripple energy-density operator the leading fermionic interaction.

\begin{figure}[t]
\centering
\begin{tikzpicture}[x=1.0cm,y=1.0cm,line cap=round,line join=round]

\begin{scope}[shift={(0,0)}]
  \draw[dirac] (-1.45,0)--(0,0);
  \draw[dirac] (0,0)--(1.45,0);
  \node[membranevertex] (V) at (0,0) {};
  \draw[philine] (0,1.25)--(V);
  \draw[philine] (V)--(0,-1.25);
  \node[panellabel] at (0,-1.65) {(a)};
\end{scope}

\begin{scope}[shift={(4.5,0)}]
  \draw[dirac] (-1.45,0)--(0,0);
  \draw[dirac] (0,0)--(1.45,0);
  \node[membranevertex] (V) at (0,0) {};
  \draw[philine] (0,1.35)--(V);
  \node[panellabel] at (0,-1.65) {(b)};
\end{scope}

\begin{scope}[shift={(9.0,0)}]
  \draw[philine] (-1.50,0)--(-0.28,0);
  \draw[philine] (0.28,0)--(1.50,0);
  \node[strainvertex,minimum width=0.48cm,minimum height=0.48cm] at (0,0) {$\times$};
  \node at (-1.22,0.42) {$\Phi_s$};
  \node at (1.22,0.42) {$\phi_e$};
  \node[panellabel] at (0,-1.65) {(c)};
\end{scope}

\end{tikzpicture}

\caption{Finite-momentum couplings.  Solid directed lines denote Dirac propagators and dashed lines denote bosonic fields.  Each arrow is centered on its uninterrupted propagator segment; open fermion lines are oriented from left to right.  Panel (a) represents the mechanical density interaction in eq.~\eqref{eq:Slambda}.  Panel (b) is the symmetry-locked Yukawa vertex in eq.~\eqref{eq:SGNY}.  Panel (c) depicts the bilinear electronic--structural mixing in eq.~\eqref{eq:LandauMix}, which induces the relation in eq.~\eqref{eq:secondary}.}
\label{fig:routes}

\end{figure}
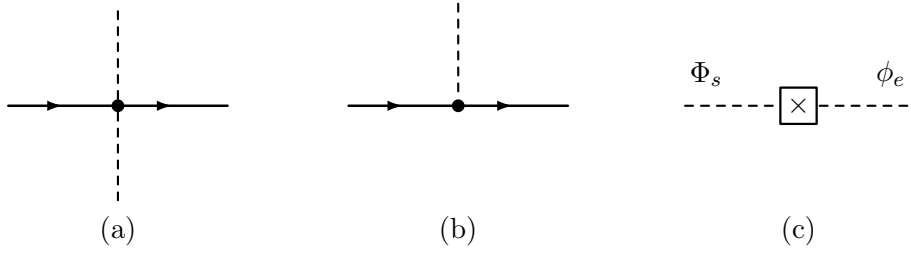

When no Dirac bilinear shares the quantum numbers of the ripple field, the leading perturbation couples the conserved fermionic density to the bosonic tuning operator,
\begin{equation}
S_\lambda=\lambda\int_{\tau,{\bf x}}|\Phi_Q|^2\rho_\psi,
\label{eq:Slambda}
\end{equation}
as represented in figure~\ref{fig:routes}(a). If particle--hole symmetry forbids this density coupling,
the leading remaining local
fermionic couplings contain additional derivatives or higher powers
of the order parameter and are therefore more irrelevant by power
counting.
At the decoupled Wilson--Fisher and free-Dirac fixed point, charge
conservation fixes the scaling dimension of $\rho_\psi$, while
$|\Phi_Q|^2$ is the tuning operator of the bosonic transition. These
relations that  the scaling dimension of the coupling $\lambda$ takes the form 
\begin{equation}
 y_\lambda=\frac{1}{\nu}-D.
\label{eq:ylambda}
\end{equation}
Expanding this relation near the Wilson--Fisher fixed point gives the one-loop result below.  For $N_b$ real components,
\begin{equation}
 \dot u=\epsilon u-{N_b+8\over6}u^2,
\end{equation}
implying  the  fixed point, 
\begin{equation}
u^*={6\epsilon\over N_b+8},
\label{eq:WFu}
\end{equation}
with $\epsilon=3-D$, and
\begin{equation}
 y_\lambda=(2-D)-{N_b+2\over N_b+8}\epsilon+O(\epsilon^2).
\label{eq:ylambda-1}
\end{equation}
For the complex order parameter describing the modes at $\pm{\bf Q}$, with two real components, the relation~\eqref{eq:ylambda} combined with the three-dimensional XY value $\nu=0.672$~\cite{Campostrini2006} gives $y_\lambda\simeq-0.511$ in $D=2$.  The one-loop $\epsilon$-expansion (eq.~\eqref{eq:ylambda-1}) gives the qualitatively consistent estimate $y_\lambda=-2/5$.  The fermions are therefore spectators at the mechanical finite-$Q$ critical point.  The corresponding one-loop Wilson--Fisher exponents are
\begin{equation}
 \nu={1\over2}+{N_b+2\over4(N_b+8)}\epsilon+O(\epsilon^2),\qquad
 \beta_\Phi={1\over2}-{3\over2(N_b+8)}\epsilon+O(\epsilon^2),
\label{eq:WFexpsMain}
\end{equation}
so that for $N_b=2$ and $\epsilon=1$ one obtains
$\nu=3/5$ and $\beta_\Phi=7/20$ at this order. These are the one-loop Wilson--Fisher critical exponents evaluated at
$\epsilon=1$.
Equations~\eqref{eq:Slambda}--\eqref{eq:WFexpsMain} follow from the symmetry, the exact scaling dimension of the conserved fermionic density, and the standard renormalization of the tuning operator.  Appendix~\ref{app:WF} gives an scaling-based argument and the one-loop $\epsilon$ expansion explicitly. When this coupling is allowed, it does not modify the leading bosonic
critical behavior. In the ordered phase, however,
$\langle\Phi_Q\rangle\neq0$, and
\begin{equation}
 \lambda |\Phi_Q|^2\rho_\psi
 \;\longrightarrow\;
 \lambda |\langle\Phi_Q\rangle|^2\rho_\psi,
 \label{eq:mueffmain}
\end{equation}
which  therefore shifts the
Dirac-point energy. At fixed chemical potential the Fermi level moves away
from the Dirac point, producing a finite carrier density; at fixed particle
number the chemical potential shifts accordingly. 

Because this coupling is irrelevant at neutrality, the structural and electronic velocities keep separate infrared values during the RG flow, in contrast to the symmetry-locked GNY route of section~\ref{sec:gny}, where a common terminal velocity and  Lorentz symmetry emerge instead.

This continuous critical theory assumes an isolated, lattice-selected pair $\pm{\bf Q}$. For an almost continuous shell of minimizing wavevectors, the classical fluctuation correction diverges as $r^{-1/2}$, whereas the shell at zero temperature  produces a logarithmic singularity. The isolated-$Q$ theory is recovered when lattice anisotropy or commensurability lifts the shell degeneracy before the critical scale is reached; see appendix~\ref{app:brazovskii}.

\section{Symmetry-locked Gross-Neveu-Yukawa route}
\label{sec:gny}

We finally turn to the case in which the finite-momentum structural field and a low-energy Dirac bilinear carry the same momentum and transform identically under the full microscopic symmetry group, including horizontal reflection.  The linear Yukawa interaction in figure~\ref{fig:routes}(b) is then allowed, and the critical order parameter contains both elastic and electronic components.  This section formulates the corresponding GNY theory \cite{OtsukaSekiSorellaYunoki2018,LiLiYao2020,LiuWangSunMeng2020}.  We give below a concrete graphene Kekul\'e mass realization and state explicitly the symmetry condition under which its coupling to a structural mode is permitted.  When that condition is satisfied, the continuum theory belongs to the
chiral-XY GNY universality class
\cite{RoyJuricicHerbut2013,FeiGiombiKlebanovTarnopolsky2016,LiJiangJianYao2017,ZerfLinMaciejko2016}.
We use its known one-loop fixed point, the correspoding  critical exponents and, for completeness, 
analyze the emergence of a common velocity for the critical bosonic
and fermionic fluctuations in the continuum limit.

\subsection{Low-energy theory}

Let $\Phi_a$ be the real components of the slowly varying structural
order parameter, and let
$\bar\psi\mathcal M_a\psi$ denote the symmetry-matched Dirac bilinear.  The Dirac nodes connected by ${\bf Q}$ are assembled into an enlarged low-energy spinor, so that the oscillating lattice phases are absorbed into the definition of the fields and the Yukawa vertex is local.  Keeping the relevant and marginal local operators near three space dimensions ($D=3-\epsilon$) gives the effective XY-GNY quantum-critical theory in the form

\begin{align}
 S_{\rm GNY}=\int d\tau\int {d^D}x\bigg\{&\bar\psi_\alpha
 \left(\gamma^0\partial_\tau+v_F\gamma^i\partial_i\right)\psi_\alpha
 +{1\over2}\left[(\partial_\tau\Phi_a)^2+c^2(\nabla\Phi_a)^2+r\Phi_a^2\right]
\nonumber\\
&+{\lambda_4\over4!}(\Phi_a^2)^2
 +g_Y\Phi_a\bar\psi_\alpha M_a\psi_\alpha\bigg\} .
\label{eq:SGNY}
\end{align}

The quartic term is the leading invariant of the $O(N_b)$ continuum theory.  Higher powers of the fields and additional derivatives are irrelevant near $d=4$.  For the commensurate graphene Kekul\'e pattern, translation symmetry permits the cubic  anisotropy
\begin{equation}
 \delta S_3=w_3\int d\tau \int {d^D}{ x}\left(\Phi^3+\Phi^{*3}\right).
\label{eq:KekuleZ3}
\end{equation}
This perturbation drives a first-order transition in the purely
bosonic Landau theory, but can become irrelevant at the
fermion-induced chiral-XY critical point. For the physical $N_f=2$
problem, sign-problem-free quantum Monte Carlo simulations and RG
analyses support the resulting emergent $U(1)$ criticality
~\cite{LiJiangJianYao2017,Roy-Juricic-PRB-Kekule}. The $O(2)$-symmetric theory below describes this infrared fixed point.  More generally, eq.~\eqref{eq:SGNY} applies to a commensurate modulation when the clock anisotropy in eq.~\eqref{eq:lockin} flows to zero.  If the clock anisotropy remains relevant, the emergent $O(2)$ fixed
point is unstable. The transition is then controlled by the
corresponding $Z_n$ theory and may become first order, as in the purely
bosonic $Z_3$ Kekul\'e problem. The velocities $v_F$ and $c$ are independent microscopic parameters, and their flow is analyzed in section~\ref{sec:velocity}.

For the mass-type channel, the Hamiltonian mass matrices
$M_a^{\rm H}$ anticommute with the Dirac kinetic Hamiltonian, so
condensation of $\Phi_a$ opens a fermion gap. A concrete spinless
graphene representation uses the valley--sublattice spinor
\begin{equation}
\Psi=\left(\psi_{K,A},\psi_{K,B},\psi_{K',A},\psi_{K',B}\right)^T,
\end{equation}
with an effective Hamiltonian
 \begin{equation}
 H_0=v_F\left(\tau_z\sigma_x k_x+\sigma_y k_y\right),
\label{eq:grapheneDiracRep}
\end{equation}
where $\tau_i$ and $\sigma_i$ act in valley and sublattice space. Physical spin supplies two identical copies of this four-component spinor, corresponding to $N_f=2$ in the RG calculation. The two Kekul\'e Hamiltonian masses at
${\bf Q}={\bf K}-{\bf K}'$ may be chosen as
$M_1^{\rm H}=\tau_x\sigma_x$ and
$M_2^{\rm H}=\tau_y\sigma_x$. They realize the two-component
chiral-XY mass algebra~\cite{HouChamonMudry2007,HerbutJuricicRoy2009},
\begin{equation}
 \{M_a^{\rm H},\tau_z\sigma_x\}
 =\{M_a^{\rm H},\sigma_y\}
 =\{M_1^{\rm H},M_2^{\rm H}\}=0.
\label{eq:KekuleMasses}
\end{equation}
Writing $\Phi=(\Phi_1+i\Phi_2)/\sqrt{2}$, the local Hamiltonian
interaction is $g_Y\Phi_a\Psi^\dagger M_a^{\rm H}\Psi$ after the
oscillating lattice phases are absorbed into the fields. The matrices appearing in eq.~\eqref{eq:SGNY} are
\begin{equation}
 \mathcal M_a=\gamma^0 M_a^{\rm H}.
\end{equation}  
The horizontal-reflection parity defines an independent selection rule.  In a free-standing mirror-symmetric sheet a pure flexural field is odd under $\sigma_h:h\mapsto-h$, whereas the usual in-plane Kekul\'e bond mass is mirror even.  Their linear Yukawa coupling is therefore forbidden even though their in-plane momentum and point-group labels match.  The graphene realization above applies when $\sigma_h$ is broken by a substrate or an asymmetric environment, or when the critical structural mode contains a mirror-even in-plane bond component with the Kekul\'e quantum numbers.  For a commensurate Kekul\'e pattern, the continuum chiral-XY description additionally requires the allowed clock anisotropy to flow to zero at criticality.  The beta functions below refer to this symmetry-allowed mass-type Clifford class; nematic bilinears define a separate anisotropic GNY theory.

The same framework describes how an electronic instability can induce a structural one, as illustrated in figure~\ref{fig:routes}(c). For a complex finite-momentum order parameter, let $\Phi_s$ and $\phi_e\sim\bar\psi M\psi$ transform in the same two-component representation.  If the electronic field becomes critical first while the structural field remains massive, the quadratic Landau functional reads

\begin{equation}
 F_2=r_s|\Phi_s|^2+r_e|\phi_e|^2
 +\kappa_{\rm mix}\left(\Phi_s^*\phi_e+\phi_e^*\Phi_s\right).
\label{eq:LandauMix}
\end{equation}
Eliminating $\Phi_s$ gives
\begin{equation}
 \Phi_s=-{\kappa_{\rm mix}\over r_s}\phi_e,
\label{eq:secondary}
\end{equation}
so an electronic condensate induces an elastic ripple. Equivalently, in real-component notation the mixing is $\kappa_{\rm mix}\Phi_{s,a}\phi_{e,a}$. Substitution back into the quadratic functional also gives
\begin{equation}
 r_e^{\rm eff}=r_e-{\kappa_{\rm mix}^2\over r_s}.
\label{eq:effectiveElectronicMass}
\end{equation}
The mixing therefore shifts the critical point and produces a soft
hybrid electronic--structural mode governed by
eq.~\eqref{eq:SGNY}.

\subsection{One-loop flow and critical exponents}

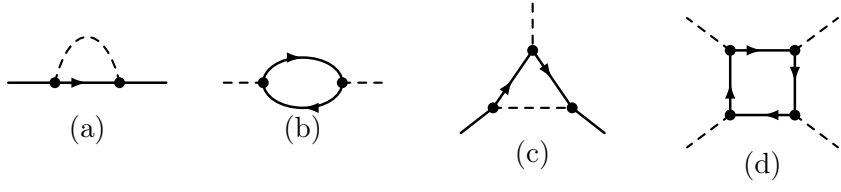
\begin{figure}[t]
\centering
\begin{tikzpicture}[scale=0.95]
\begin{scope}[shift={(0,0)}]
\draw[diracplain] (-1.1,0)--(-0.45,0);
\draw[dirac] (-0.45,0)--(0.45,0);
\draw[diracplain] (0.45,0)--(1.1,0);
\draw[boson] (-0.45,0) .. controls (-0.25,0.85) and (0.25,0.85) .. (0.45,0);
\node[vertex] at (-0.45,0) {};
\node[vertex] at (0.45,0) {};
\node at (0,-0.65) {(a)};
\end{scope}
\begin{scope}[shift={(3.0,0)}]
\draw[philine] (-1.1,0)--(-0.55,0);
\draw[philine] (0.55,0)--(1.1,0);
\draw[dirac] (-0.55,0) arc[start angle=180,end angle=0,x radius=0.55,y radius=0.35];
\draw[dirac] (0.55,0) arc[start angle=0,end angle=-180,x radius=0.55,y radius=0.35];
\node[vertex] at (-0.55,0) {};
\node[vertex] at (0.55,0) {};
\node at (0,-0.65) {(b)};
\end{scope}
\begin{scope}[shift={(6.2,0)}]
\node[vertex] at (0,0.45) {};
\node[vertex] at (-0.55,-0.35) {};
\node[vertex] at (0.55,-0.35) {};
\draw[philine] (0,1.1)--(0,0.45);
\draw[diracplain] (-1.0,-0.7)--(-0.55,-0.35);
\draw[dirac] (-0.55,-0.35)--(0,0.45);
\draw[dirac] (0,0.45)--(0.55,-0.35);
\draw[diracplain] (0.55,-0.35)--(1.0,-0.7);
\draw[boson] (-0.55,-0.35)--(0.55,-0.35);
\node at (0,-1.0) {(c)};
\end{scope}
\begin{scope}[shift={(9.4,0)}]
\coordinate (a) at (-0.45,0.45);
\coordinate (b) at (0.45,0.45);
\coordinate (c) at (0.45,-0.45);
\coordinate (d) at (-0.45,-0.45);
\draw[dirac] (a)--(b);
\draw[dirac] (b)--(c);
\draw[dirac] (c)--(d);
\draw[dirac] (d)--(a);
\node[vertex] at (a) {};
\node[vertex] at (b) {};
\node[vertex] at (c) {};
\node[vertex] at (d) {};
\draw[philine] (a)--(-1.05,0.9);
\draw[philine] (d)--(-1.05,-0.9);
\draw[philine] (b)--(1.05,0.9);
\draw[philine] (c)--(1.05,-0.9);
\node at (0,-1.15) {(d)};
\end{scope}
\end{tikzpicture}

\caption{Representative one-loop diagrams of the finite-$Q$ GNY theory. Solid lines denote Dirac propagators and dashed lines denote bosonic order-parameter propagators. Arrows are centered on the internal fermion segments between vertices; open fermion lines are oriented from left to right, while arrows on closed loops follow a consistent circulation. Panels (a) and (b) determine the fermionic and bosonic anomalous dimensions in eq.~\eqref{eq:XY-anomalous-dimensions}; panels (c) and (d) contribute to the Yukawa and quartic beta functions in eqs.~\eqref{eq:betax} and~\eqref{eq:betau}. The pure bosonic quartic bubble and the bosonic mass-insertion diagram are not shown. The pole parts and normalization conventions are collected in appendix~\ref{app:gny}.}
\label{fig:gnygraphs}

\end{figure}

We analyze the GNY theory in $d=4-\epsilon$ spacetime dimensions using four-component Dirac fermions, $\bluerev{\Tr\,\mathbf{1}=N_D=4}$. For the fixed-point couplings and critical exponents, we rescale Euclidean time and the fields so that the reference boson velocity is one and evaluate the dimensionless couplings at the Lorentz-symmetric (equal-velocity) point $v_F=c=1$. Section~\ref{sec:velocity} then shows that this point is stable within the isotropic continuum theory. With $\lambda_4$ denoting the dimensionful quartic coefficient in eq.~\eqref{eq:SGNY}, we define
\begin{equation}
 x={\mu_{\rm RG}^{-\epsilon}g_Y^2\over4\pi^2},
 \qquad
 u={\mu_{\rm RG}^{-\epsilon}\lambda_4\over8\pi^2}.
\label{eq:hxdef}
\end{equation}
In this convention, the one-loop beta functions and anomalous dimensions for the mass-type Clifford class are
\begin{align}
 \dot x&=\epsilon x-A_hx^2,
\label{eq:betax}\\
 \dot u&=\epsilon u-A_uu^2-4N_fxu+24N_fx^2,
\label{eq:betau}
\end{align}
where $A_h=2N_f+4-N_b$, $A_u=(N_b+8)/6$, and the bosonic and
fermionic anomalous dimensions are $\eta_\Phi=2N_fx$ and
$\eta_\psi=N_bx/2$, respectively. Figure~\ref{fig:gnygraphs} displays
representative one-loop diagrams; the pure bosonic quartic bubble and
the bosonic mass-insertion diagram are not shown. Appendix~\ref{app:gny}
collects the singular parts and normalization conventions.

The interacting fixed point is

\begin{equation}
 x^*={\epsilon\over A_h},
 \qquad
 u^*=\epsilon\bar u_*,
\label{eq:GNYfp}
\end{equation}

where $\bar u_*$ is the positive solution of

\begin{equation}
 0=\bar u_*-A_u\bar u_*^2-{4N_f\over A_h}\bar u_*
 +{24N_f\over A_h^2}.
\label{eq:ubarstar_equation}
\end{equation}
For the symmetry-allowed graphene Kekul\'e chiral-XY case, $N_f=2$ and $N_b=2$, so $A_h=6$, $x^*=\epsilon/6$, $\bar u_*=4/5$, and $u^*=4\epsilon/5$. The correlation length is governed by the scaling dimension $y_r$ of the bosonic tuning parameter. Writing $\dot r=y_r r$, one finds
\begin{equation}
 y_r=2-{N_b+2\over6}u-2N_fx,
\label{eq:rflow}
\end{equation}
At the interacting fixed point, the correlation-length exponent is therefore determined by
\begin{equation}
 {1\over\nu}=y_r^*
 =2-{N_b+2\over6}u^*-2N_fx^*.
\label{eq:invnu}
\end{equation}
For $N_f=N_b=2$, the interacting fixed point yields
\begin{equation}
 \nu=\frac{1}{2}+\frac{3}{10}\epsilon+O(\epsilon^2),
 \label{eq:grapheneGNY}
\end{equation}
yielding the one-loop $\epsilon$-expansion result $\nu=4/5$ at $\epsilon=1$. The anomalous dimensions are
\begin{equation}
 \eta_\Phi=\frac{2}{3}\epsilon+O(\epsilon^2),
 \qquad
 \eta_\psi=\frac{1}{6}\epsilon+O(\epsilon^2).
 \label{eq:XY-anomalous-dimensions}
\end{equation}
These results agree with the one-loop chiral-XY GNY coefficients in refs.~\cite{RoyJuricicHerbut2013,Roy-Juricic-PRB2014,FeiGiombiKlebanovTarnopolsky2016,ZerfLinMaciejko2016}.  Complementary large-$N$ and higher-loop estimates are available in refs.~\cite{Gracey2021ChiralXY,GraceyMaierMarquardSchroder2025}.

\subsection{Emergent terminal velocity in the isotropic continuum theory}
\label{sec:velocity}

Let
\begin{equation}
 \zeta={v_F\over c}
\label{eq:zetadef}
\end{equation}
be the velocity ratio. After rescaling Euclidean time by the reference boson velocity, we define the  dimensionless coupling at the Lorentz-symmetric (equal-velocity) critical point by
\begin{equation}
 \widehat g^{\,2}={x\over2}.
\label{eq:g2dict}
\end{equation}
All smooth velocity-dependent prefactors are evaluated at $\zeta=1$, while the factor $(\zeta-1)$ controls the linearized flow.  We then find near $\zeta=1$

\begin{align}
 {\dd\ln v_F\over\dd\ell}&=-{N_b\over6}\widehat g^{\,2}(\zeta-1)+O((\zeta-1)^2),
\label{eq:fermvelmain}\\
 {\dd\ln c\over\dd\ell}&=N_f\widehat g^{\,2}(\zeta-1)+O((\zeta-1)^2),
\label{eq:bosvelmain}
\end{align}
and therefore
\begin{equation}
 {\dd\zeta\over\dd\ell}=-\left(N_f+{N_b\over6}\right)
 \widehat g^{\,2}\zeta(\zeta-1)+O((\zeta-1)^2).
\label{eq:zetaflow}
\end{equation}
Linearizing eq.~\eqref{eq:zetaflow} about $\zeta=1$ gives
\begin{equation}
 \frac{\dd(\zeta-1)}{\dd\ell}
 =-\omega_\zeta(\zeta-1),
 \label{eq:omega-zeta}
\end{equation}
where 
\begin{equation}
 \omega_\zeta=
 \left(N_f+\frac{N_b}{6}\right)\widehat g_*^{\,2}.
\end{equation}
For $N_f=N_b=2$ and
$\widehat g_*^{\,2}=\epsilon/12$, one obtains
\begin{equation}
 \omega_\zeta=\frac{7}{36}\epsilon.
\end{equation}
The equal-velocity point therefore stable in the isotropic
mass-type GNY theory~\cite{RoyHerbut2016}. A generic finite-$Q$ state permits independent
$c_\parallel,c_\perp,v_{F,\parallel},v_{F,\perp}$ which are expected to flow to a common terminal velocity. The counterterms
underlying these flows are collected in appendix~\ref{app:gny}.

\section{Conclusions and outlook}
\label{sec:conclusions}

We have developed a low-energy field theory for  Dirac fermions coupled to crystalline membranes, covering both the asymptotic flat phase and finite-momentum rippling instabilities.  For the leading scalar projection of the local interaction~\eqref{eq:Sg}, the isotropic-tension Ward identity determines the scaling of the flexural-strain operator $H_T$, while fermionic corrections are suppressed by the infrared-irrelevant coupling $\alpha_e$.  Consequently, the logarithmic stiffening of the membrane persists and the membrane remains stable against this perturbation: the dimensionless strength of the leading local scalar interaction decreases logarithmically at long wavelengths, the induced nonlocal quartic strain interaction vanishes as a power law, and the fermion polarization leaves the flexural dynamics undamped.

Finite-momentum rippling instabilities define distinct relativistic effective theories once microscopic physics produces isolated minima at $\pm{\bf Q}$. When no low-energy Dirac bilinear carries the  quantum numbers of the ripple field, including the momentum, the leading allowed interaction couples the conserved fermion density to the bosonic tuning operator. The scaling dimension of this fermion--boson coupling is $y_\lambda=1/\nu-D$. For the two real components of the complex ripple order parameter in $D=2$, one has $y_\lambda<0$, and the coupling is irrelevant. The transition is governed by the purely bosonic $\Phi^4$ Wilson--Fisher fixed point, while the Dirac fermions remain spectators at criticality. The ordered phase can nevertheless shift the Dirac point relative to the chemical potential. A relevant clock anisotropy destabilizes the XY fixed point, while an approximately continuous ring of minima leads to the Brazovskii regime discussed above.

A qualitatively different critical theory arises when the structural field and a mass-type Dirac bilinear transform identically under the full microscopic symmetry group.  The critical mode is then a hybrid electronic--structural order parameter governed, in the simplest two-component realization, by the chiral-XY Gross--Neveu--Yukawa universality class.  In graphene, the two valley-mixing Kekul\'e bilinears provide the mass components required by the chiral-XY theory.  A linear coupling to a pure flexural ripple, however, requires broken horizontal-reflection symmetry or a critical distortion with a symmetry-compatible bond component, since the flexural ripple is mirror odd whereas the Kekul\'e mass is mirror even.  Under these conditions, the known chiral-XY critical exponents characterize the transition, an electronic instability induces a secondary structural distortion, and the fermionic and bosonic velocities flow toward a common value at criticality within the isotropic continuum theory.

This framework can be extended to the full momentum-projected
deformation potential and to the independent shear-channel
pseudogauge coupling to the valley current~\cite{VozmedianoKatsnelsonGuinea2010,Manes2007,GuineaHorovitzLeDoussal2008,GuineaKatsnelsonGeim2010}.
Further directions include commensurate lock-in perturbations,
multi-$Q$ patterns, fermionic nematic instabilities, and finite-density
regimes in which the polarization can modify the dynamical
universality class.

\begin{acknowledgments} 
We thank Bitan Roy for the critical reading of the manuscript. This work is supported by Fondecyt (Chile) Grants  No. 1230933 (V.J.), 1230440 (E.M.) and 1241033 (R.S.-G.). 
\end{acknowledgments}

\appendix

\section{Feynman rules and diagrammatic conventions}
\label{app:feynman}

We collect here the propagators and diagrammatic conventions used below.
The flexural propagator is

\begin{equation}
 G_h(\omega,{\bf k})={1\over \rho \omega^2+\kappa_bk^4}.
\label{eq:Gh}
\end{equation}

The Dirac propagator at charge neutrality is

\begin{equation}
 G_\psi(i\omega,{\bf k})={-i\omega\gamma^0+v_F k_i\gamma^i\over \omega^2+v_F^2 k^2}.
\label{eq:Gpsi}
\end{equation}
The finite-$Q$ order-parameter propagator at criticality is
\begin{equation}
 G_\Phi(\Omega,{\bf q})={1\over \Omega^2+c^2q^2}.
\label{eq:Gphi}
\end{equation}
Wiggly lines denote long-wavelength flexural propagators, dashed scalar lines denote finite-$Q$ order-parameter propagators, and solid directed lines denote Dirac fermions, see figure~\ref{fig:flatgraphs}. The local strain insertion $\HT$ carries two height legs and a factor proportional to $(k_1\cdot k_2)$ projected onto the trace channel.

\section{Pure membrane and the tension Ward identity}
\label{app:strain}

The composite strain relevant for the scalar deformation potential is

\begin{equation}
 \HT={1\over2}(\partial_i h_a)^2,
\end{equation}

An isotropic tension couples to this operator through

\begin{equation}
 S_\sigma={\sigma\over2}\int\dd\tau\dd^Dx\,(\partial_i h_a)^2.
\label{eq:StensionApp}
\end{equation}

The exact nonlinear symmetry is most directly expressed through the complete induced-metric strain,
\begin{equation}
 u_{ij}=\frac12\left(
 \partial_i u_j+\partial_j u_i
 +\partial_i u_k\partial_j u_k
 +\partial_i h_a\partial_j h_a
 \right).
\label{eq:strainOriginUij}
\end{equation}
For the embedding position $R_\mu(x)=(x_i+u_i(x),h_a(x))$, an infinitesimal rotation mixing the in-plane direction $i$ and the transverse direction $a$ acts in Monge gauge as~\cite{AronovitzGolubovicLubensky1989}
\begin{equation}
 \delta u_i=\Xi_{ia}h_a(x),
 \qquad
 \delta h_a=-\Xi_{ia}\left[x_i+u_i(x)\right].
\label{eq:strainOriginTransf}
\end{equation}
The complete elastic theory constructed from the induced metric obeys the corresponding nonlinear rotational Ward identity~\cite{GuitterDavidLeiblerPeliti1989}. Expanding this identity to the order retained in the long-wavelength action gives the relation required here: at vanishing external tension, no ultraviolet-divergent $p^2$ term is generated in the flexural two-point function. The divergent self-energy therefore starts at order $p^4$. Appendix~\ref{app:quantum} verifies this consequence explicitly at one loop. We use only this consequence of the full Ward identity; its complete functional derivation is not reproduced here.

Accordingly, at zero external frequency the divergent momentum-dependent part of the flexural self-energy has the form
\begin{equation}
 \Sigma_h^{\rm div}(0,{\bf p})=A_h p^4,
\label{eq:selfenergyfactor}
\end{equation}
so the divergent self-energy begins at order $p^4$.  Differentiating the inverse propagator with respect to $\sigma$ generates the zero-momentum insertion of $\HT$,

\begin{equation}
 \Gamma_{H}^{(2)}(p)={\partial\Gamma_h^{(2)}(p;\sigma)\over\partial\sigma}\bigg|_{\sigma=0}.
\label{eq:insertionDerivative}
\end{equation}

Equation~\eqref{eq:selfenergyfactor} therefore fixes the scaling of $\Gamma_H^{(2)}$ entirely through the height-field and bending-rigidity renormalizations. Hence
\begin{equation}
 y_\sigma=2-\eta_h.
\label{eq:WardygApp}
\end{equation}
Combining this static tension exponent with the scaling dimension $\Delta_\rho=D$ of the conserved density gives $y_g^{\rm cl}=2-D-\eta_h$. No additional vertex factor appears in the zero-momentum tension channel. At finite momentum transfer, derivative operators may mix with $\HT$ and require a separate composite-operator analysis. Additive mixing of $\HT$ with the identity shifts only the background tension and does not modify the tension scaling.

\section{Dirac polarization}
\label{app:polarization}

The density bubble is
\begin{equation}
 \Pi_{00}(i\Omega,{\bf q})=-N_f\int_{\omega,{\bf k}}
 \Tr\left[\gamma^0G_\psi(i\omega+i\Omega,{\bf k}+{\bf q})
 \gamma^0G_\psi(i\omega,{\bf k})\right].
\label{eq:PiIntegral}
\end{equation}
Using $\Tr\,\mathbf{1}=N_D$ and performing the standard Feynman-parameter integral in $(2+1)$ dimensions gives
\begin{equation}
 \Pi_{00}(i\Omega,{\bf q})
 =-{\calN\over16}{q^2\over\sqrt{v_F^2q^2+\Omega^2}},
\label{eq:PiAppResult}
\end{equation}
with $\calN=N_fN_D/2$.  At charge neutrality, the polarization contains no term proportional to $|\Omega|/|q|$, because the Dirac system has no Fermi surface and therefore does not exhibit Landau damping.  At finite chemical potential, the polarization crosses over to the metallic Hertz-Millis form in the small-$q$ regime.

\section{Quantum composite operator and controlled mixed flow}
\label{app:quantum}

Before evaluating the insertion, it is useful to make the normalization of the mixed coupling explicit.  Starting from the quadratic flexural kernel $\rho\omega^2+\kappa_b k^4$, set

\begin{equation}
 \tau=\sqrt{\rho/\kappa_b}\,\widetilde\tau,
 \qquad
 h_a=(\rho\kappa_b)^{-1/4}\widetilde h_a.
\label{eq:canonicalRescaling}
\end{equation}

The flexural quadratic action is then canonical, while

\begin{equation}
 \widetilde v_F=v_F\sqrt{\rho/\kappa_b},
 \qquad
 \bar g={g\over\kappa_b},
 \qquad
 s={k\over\widetilde v_F}.
\label{eq:canonicalMixedParams}
\end{equation}
Thus the local running vertex is the microscopic deformation potential measured in units of the scale-dependent bending coefficient.  The Ward identity below fixes its entire pure-membrane renormalization.

In this appendix we denote the dimensional regulator by
$\epsilon_q=2-D$, distinguishing it from the relativistic regulator
$\epsilon=3-D$ used in the finite-$Q$ sectors. To isolate the logarithmic
ultraviolet divergence, we analytically continue the loop-momentum
integrals to $D=2-\epsilon_q$ while retaining the two-dimensional
tensor structure of the vertex.

At the physical spatial dimension $D=2$, introducing the
Hubbard--Stratonovich field $\chi$ brings the action to the form
\begin{equation}
 S_h=\int\dd\tau\dd^2x\left[
 {1\over2}\dot h_a^2+{1\over2}(\nabla^2h_a)^2
 +{\sigma\over2}(\partial_i h_a)^2
 +{1\over2y}(\nabla^2\chi)^2+i\chi\mathcal K[h]
 \right],
\label{eq:auxAction}
\end{equation}
where
\begin{equation}
 \mathcal K[h]={1\over2}\left[(\nabla^2h_a)^2-
 (\partial_i\partial_jh_a)^2\right].
\label{eq:GaussianCurv}
\end{equation}
The three-field vertex is
\begin{equation}
 \gamma({\bf k}_1,{\bf k}_2)=k_1^2k_2^2-({\bf k}_1\cdot{\bf k}_2)^2.
\label{eq:curvvertex}
\end{equation}
It contains two powers of the momentum of each external height field.  The one-loop height self-energy therefore behaves for small external momentum as
\begin{equation}
 \Sigma_h^{(1)}(0,{\bf p})
 \sim y p^4\int{\dd\omega\dd^{2-\epsilon_q}k\over(2\pi)^{3-\epsilon_q}}
 {\sin^4\vartheta\over\omega^2+k^4}
 \propto {y\over\epsilon_q}p^4,
\label{eq:quantumSelfSchematic}
\end{equation}
which produces the bending-rigidity counterterm while leaving the $p^2$ coefficient protected.  An insertion of $\HT$ at zero momentum is obtained by differentiating an internal height propagator with respect to $\sigma$,

\begin{equation}
 {\partial\over\partial\sigma}
 {1\over\omega^2+k^4+\sigma k^2}\bigg|_{\sigma=0}
 =-{k^2\over(\omega^2+k^4)^2}.
\label{eq:tensionInsertionProp}
\end{equation}
After the frequency integration, the large-momentum behavior of the corresponding one-loop insertion is proportional to
\begin{equation}
 y p^4\int^{\Lambda}{\dd^{2-\epsilon_q}k\over(2\pi)^{2-\epsilon_q}}{1\over k^4},
\label{eq:insertionUVfinite}
\end{equation}
which is ultraviolet convergent at $D=2$ and infrared singular when all external scales are set to zero. A nonzero external momentum, frequency, or tension is therefore retained only as an infrared regulator. Since no ultraviolet pole is generated, the local tension insertion requires no independent vertex counterterm. Thus
\begin{equation}
 Z_H^{\rm(vtx)}=1.
\label{eq:ZHquantum}
\end{equation}
The full rotational Ward identity ensures that the zero-momentum
tension insertion has no independent renormalization beyond the
height-field and bending-rigidity factors.  In the one-loop normalization used in the main text, the membrane renormalizations are
\begin{equation}
 {\dd y\over\dd\ell}=-{3(d_c+6)\over128\pi}y^2,
 \qquad
 {\dd\ln\kappa_b\over\dd\ell}={3\over32\pi}y.
\label{eq:qRGapp}
\end{equation}

To control the fermionic feedback on slow flexural observables, we evaluate the fermionic determinant at $\omega\sim\sqrt{\kappa_b/\rho}\,k^2$.  The Dirac propagator admits the uniform expansion
\begin{equation}
 G_\psi(i\omega,{\bf k})=
 {\gamma^i\widehat k_i\over v_F k}
 +O\left({s\over v_F k}\right),
 \qquad
 s={\omega\over v_Fk}=O\left({\sqrt{\kappa_b/\rho}\,k\over v_F}\right).
\label{eq:fastFermionExpansion}
\end{equation}
The Dirac density--density polarization function at neutrality (zero chemical potential) 
calculated in appendix~\ref{app:polarization} has a leading static
term proportional to $|q|/v_F$. Multiplication by the two normalized
vertices gives the coefficient
$\alpha_e=({\cal N}/16)\bar g^{\,2}s$. Higher terms in the derivative expansion carry additional powers of $s$.  Since
\begin{equation}
 s(\ell)\sim e^{-\ell}\ell^{\theta/2},
 \qquad
 \alpha_e(\ell)\sim e^{-\ell}\ell^{-3\theta/2},
\label{eq:mixedAsymptoticsApp}
\end{equation}
the accumulated fermion-induced elastic corrections remain finite during  the infrared flow and subleading to the logarithmic pure-membrane flow.

We next verify the power counting of the mixed 1PI vertex. After extracting the bare derivative factor, we decompose the projected vertex as
\begin{equation}
 \Gamma^{\rm 1PI}_{hh\bar\psi\psi}
 =({\bf p}_1\!\cdot{\bf p}_2)\gamma^0
 \left[g_R+\delta g_R(P)\right]+\Gamma_{\rm der},
\label{eq:bareMixedVertex}
\end{equation}
where $\Gamma_{\rm der}$ contains operators with additional external derivatives. The order-$g^2$ diagram is
\begin{equation}
 \delta\Gamma^{(2)}_{hh\bar\psi\psi}
 \sim g^2\!\int_{\omega,{\bf k}}
 ({\bf p}_1\!\cdot{\bf k})({\bf p}_2\!\cdot{\bf k})
 G_h(i\omega,{\bf k})\,\gamma^0G_\psi(i\omega,{\bf k})\gamma^0.
\label{eq:mixedVertexOneLoop}
\end{equation}
Retaining a small external fermion momentum $P$, we expand $G_\psi(K-P)=G_\psi(K)-P_\mu\partial_{K_\mu}G_\psi(K)+O(P^2)$. The $P^0$ term is odd under $(\omega,{\bf k})\mapsto(-\omega,-{\bf k})$ and vanishes. The first surviving term is proportional to $P_\mu$ and belongs to $\Gamma_{\rm der}$, so the one-loop diagram does not renormalize $g_R$. The first possible correction to the original local tensor structure
contains two additional density--strain vertices and is of order
$g^3$. Relative to the tree-level vertex, the additional pair of
normalized density--strain vertices supplies $\bar g^{\,2}$, while
the slow-flexural expansion of the fast Dirac denominator supplies
at least one factor $\omega_h/(v_Fk)=s$. Thus the projected local
correction satisfies the bound
\begin{equation}
 \frac{\delta g_R}{g_R}
 =O\!\left(\bar g^{\,2}s\right),
\end{equation}
with possible additional powers of $s$. In the slow-flexural regime, power counting then gives
\begin{equation}
 \gamma_g^{(\psi)}
 =O(\bar g^{\,2}s)
 =O(\alpha_e),
\label{eq:mixedVertexPowerApp}
\end{equation}
up to finite numerical and flavor-dependent factors. 
In the fast regime  $\omega\sim v_Fk$, the flexural propagator is
controlled by its inertial term and scales as $G_h\sim k^{-2}$ rather
than $k^{-4}$. Each internal flexural line is therefore less infrared
singular than under the slow scaling $\omega\sim k^2$, so the fast
region cannot generate an $s^0$ logarithm in the projected local
vertex. Hence no $s^0$ logarithm modifies the Ward-identity scaling, and the fermionic contribution to $\dd\ln\bar g/\dd\ell$ vanishes with $\alpha_e$.

The complementary check concerns fermionic observables, for which $\Omega\sim v_Fq$.  At Gaussian order the connected trace-strain correlator is
\begin{equation}
 \mathcal D_H(i\Omega,{\bf q})
 ={d_c\over2}\int_{\omega,{\bf k}}
 \left[{\bf k}\!\cdot({\bf q}-{\bf k})\right]^2
 G_h(i\omega,{\bf k})G_h(i\Omega-i\omega,{\bf q}-{\bf k}).
\label{eq:strainCorrelatorConvolution}
\end{equation}
At $D=2$ and $z=2$, rescaling ${\bf k}=q{\bf y}$ and $\omega=C_Hq^2 z$, with $C_H=\sqrt{\kappa_b/\rho}$, gives the full scaling form
\begin{equation}
 \mathcal D_H(i\Omega,{\bf q})
 ={d_c\over\rho^{1/2}\kappa_b^{3/2}}
 \mathcal F\!\left({\Omega\over C_H q^2},{\Lambda\over q}\right).
\label{eq:strainCorrelatorScaling}
\end{equation}
For $|\Omega|\ll C_Hq^2$, the frequency integral leaves $\int^{\Lambda}\dd^2k/[k^2({\bf q}-{\bf k})^2]$ multiplied by the derivative numerator, producing at most $\ln(\Lambda/q)$.  For $|\Omega|\gg C_Hq^2$, the frequency supplies the infrared scale $\sqrt{|\Omega|/C_H}$ and the same argument gives at most $\ln[\Lambda\sqrt{C_H/|\Omega|}]$.  Equivalently, with
\begin{equation}
 \mu_Q=\max\!\left(q,\sqrt{|\Omega|/C_H}\right),
\end{equation}
one may use the bound
\begin{equation}
 |\mathcal D_H(i\Omega,{\bf q})|
 \lesssim {d_c\over\rho^{1/2}\kappa_b^{3/2}}
 \left[1+\ln{\Lambda\over\mu_Q}\right],
\label{eq:strainCorrelatorBound}
\end{equation}
where the marginal elastic interaction changes only the logarithmic factors.  This statement uses the full $({\Omega},{\bf q})$ dependence rather than the special ${\bf q}=0$ limit.

The corresponding fermion self-energy is
\begin{equation}
 \Sigma_\psi(P)=g^2\int_Q
 \mathcal D_H(Q)\gamma^0G_\psi(P-Q)\gamma^0.
\label{eq:fermionSelfEnergyMixed}
\end{equation}

In the infrared region $Q\sim P$ of the relativistic fermion kinematics, we denote by $p=\sqrt{P_0^2+v_F^2{\bf P}^2}/v_F$ the corresponding momentum scale. The integration measure scales as $p^3$, the Dirac propagator as $p^{-1}$, and eq.~\eqref{eq:strainCorrelatorBound} contributes at most a logarithm. After the analytic local $P^0$ and $P^1$ contributions from the ultraviolet region are absorbed into local parameter renormalizations, the nonanalytic infrared remainder obeys

\begin{equation}
 \Sigma_\psi^{\rm IR}(P)
 =O\left({g^2\over\rho^{1/2}\kappa_b^{3/2}}p^2
 \ln{\Lambda\over p}\right),
\label{eq:fermionSelfScaling}
\end{equation}
with no $p\ln p$ term. The analytic terms through first order in $P$
renormalize local short-distance parameters, while the remaining
nonanalytic contribution is subleading to the Dirac kinetic term.
Neither the fermion field nor $v_F$ therefore acquires an infrared
logarithmic anomalous dimension from the flexural sector. Equations~\eqref{eq:mixedAsymptoticsApp}, \eqref{eq:mixedVertexPowerApp}, and \eqref{eq:fermionSelfScaling} establish the controlled expansion in the mixed flexural-Dirac sector used in section~\ref{sec:Tzero}.

\section{Mechanical finite-\texorpdfstring{$Q$}{Q} criticality}
\label{app:WF}

For this channel, the leading scalar perturbation is given by  eq.~\eqref{eq:Slambda}.   The charge density is the temporal component of a conserved current and therefore has the exact dimension $\Delta_\rho=d-1=D$, while the scaling dimension of bosonic tuning operator is $\Delta_{\Phi^2}=d-1/\nu$. This therefore yields for the scaling dimension of the coupling 
\begin{equation}
 y_\lambda=d-\Delta_\rho-\Delta_{\Phi^2}
 ={1\over\nu}-D,
\label{eq:ylambdaAppExact}
\end{equation}
and implies eq.~\eqref{eq:ylambda} independently of perturbation theory.

For $N_b$ real components the relativistic order-parameter theory is
\begin{equation}
 S_\Phi=\int_x\left[{1\over2}(\partial_\mu\Phi_a)^2+{r\over2}\Phi_a^2+{\lambda_4\over4!}(\Phi_a^2)^2\right].
\label{eq:Ophi}
\end{equation}
We define the dimensionless coupling
\begin{equation}
 u={\mu_{\rm RG}^{-\epsilon}\lambda_4\over8\pi^2}.
\label{eq:WFcouplingDefApp}
\end{equation}

Following the standard critical-phenomena treatment of the $O(N_b)$ model \cite{ZinnJustin2002}, the one-loop RG flows read
\begin{equation}
 \dot u=\epsilon u-{N_b+8\over6}u^2,
 \qquad
 {1\over\nu}=2-{N_b+2\over6}u^*.
\label{eq:WFapp}
\end{equation}
Expanding the inverse before setting $\epsilon=1$ gives
$\nu=1/2+(N_b+2)\epsilon/[4(N_b+8)]+O(\epsilon^2)$.  Substitution into eq.~\eqref{eq:ylambdaAppExact} gives the one-loop
expansion in eq.~\eqref{eq:ylambda-1}.  Since $\eta_\Phi=O(\epsilon^2)$ at the Wilson--Fisher point, the order-parameter exponent follows from $\beta_\Phi=\nu(d-2+\eta_\Phi)/2$, yielding the second relation in eq.~\eqref{eq:WFexpsMain}.  For $N_b=2$ the corresponding one-loop values are therefore $\nu=3/5$ and $\beta_\Phi=7/20$.

Using instead the estimate for three-dimensional XY model $\nu=0.672$~\cite{Campostrini2006} gives the nonperturbative numerical value $y_\lambda\simeq-0.511$.  In the ordered phase, the condensate contribution shifts the Dirac-point energy by an amount proportional to $-\lambda_R|\langle\Phi_Q\rangle|^2$, as described in eq.~\eqref{eq:mueffmain}.

\section{GNY one-loop calculation}
\label{app:gny}

The fermion self-energy is
\begin{equation}
 \Sigma_\psi(p)=g_Y^2\sum_{a=1}^{N_b}\int_k \mathcal M_a G_\psi(k)\mathcal M_a G_\Phi(p-k).
\label{eq:SigmaPsiApp}
\end{equation}

It contains no closed fermion trace and is proportional to $N_b$. In the four-component convention
$x=\mu_{\rm RG}^{-\epsilon}g_Y^2/(4\pi^2)$, projection onto the kinetic term gives

\begin{equation}
 \eta_\psi={N_b\over2}x.
\label{eq:etaPsiApp}
\end{equation}
This is the same four-component normalization used in section~\ref{sec:gny}.

The boson self-energy is
\begin{equation}
 \Sigma_{\Phi,ab}(p)=-g_Y^2N_f\int_k\Tr\left[\mathcal 
 M_a G_\psi(k+p)\mathcal M_b G_\psi(k)\right],
\label{eq:SigmaPhiApp}
\end{equation}
and carries the four-component Clifford trace $\Tr(\mathcal M_a \mathcal M_b)=4\delta_{ab}$, giving the bosonic anomalous dimension
\begin{equation}
 \eta_\Phi=2N_fx.
\label{eq:etaPhiApp}
\end{equation}
We here use $N_D=4$ and the definitions
\[
 x={\mu_{\rm RG}^{-\epsilon}g_Y^2\over4\pi^2},
 \qquad
 u={\mu_{\rm RG}^{-\epsilon}\lambda_4\over8\pi^2},
\]
while the pole counterterms are defined  through the corresponding
bare--renormalized relations for $x$, $u$, and $r$; only their
minimal-subtraction pole parts are displayed below.

In the minimal-subtraction scheme, the one-loop anomalous dimensions
and pole counterterms are
\begin{align}
 \eta_\psi&=\frac{N_b}{2}x,
 &\eta_\Phi&=2N_f x,
 \notag\\
 \delta x_{\rm pole}
 &=\frac{A_h}{\epsilon}x^2,
 &A_h&=2N_f+4-N_b,
 \notag\\
 \delta u_{\rm pole}
 &=\frac{1}{\epsilon}
 \left(A_u u^2+4N_fxu-24N_fx^2\right),
 &A_u&=\frac{N_b+8}{6}.
 \label{eq:GNYpoleSummary}
\end{align}
The three contributions to $A_h$ are $2N_f$ from the boson
wave-function renormalization, $N_b$ from the fermion wave-function
renormalization, and $4-2N_b$ from the Yukawa-vertex renormalization.
With the infrared RG convention adopted in the main text, these pole
terms yield eqs.~\eqref{eq:betax} and~\eqref{eq:betau}.

The bosonic tuning parameter is renormalized by the zero-momentum
two-point function with one mass insertion. In the normalization of
eq.~\eqref{eq:hxdef}, its one-loop pole counterterm is
\begin{equation}
 \delta r_{\rm pole}
 =
 \frac{r}{\epsilon}
 \left[
 \frac{N_b+2}{6}u+2N_f x
 \right].
\end{equation}
With the same infrared RG convention, this gives
\begin{equation}
 y_r
 =
 2-\frac{N_b+2}{6}u-2N_f x,
\end{equation}
reproducing eq.~\eqref{eq:rflow}.

The velocity flow is obtained from the difference between the fermionic (bosonic) temporal kinetic  and spatial kinetic  counterterms, respectively, $\delta Z_{\psi,\tau}$ and $\delta Z_{\psi,x}$ ($\delta Z_{\Phi,\tau}$ and $\delta Z_{\Phi,x}$).  We then define the dimensionless coupling $\widehat g^{\,2}=(\mu_{\rm RG}^{-\epsilon}g_Y^2)/(8\pi^2)$. Only its expansion about the Lorentz-symmetric (equal-velocity) point is needed below; in the convention $c=1$ and at $v_F=c$, it reduces to $\widehat g^{\,2}=x/2$.  From the fermion and boson self-energies we obtain
\begin{equation}
 \delta Z_{\psi,\tau}-\delta Z_{\psi,x}
 ={N_b\over6}\widehat g^{\,2}(\zeta-1)\frac{1}{\epsilon},
 \qquad
 \delta Z_{\Phi,\tau}-\delta Z_{\Phi,x}
 =-N_f\widehat g^{\,2}(\zeta-1)\frac{1}{\epsilon},
\label{eq:velocityPoleDifferences}
\end{equation}
where only the difference is required for the velocity ratio.  Near $\zeta=v_F/c=1$, we find  the  RG flows of the form 
\begin{align}
 {\dd\ln v_F\over\dd\ell}&=-{N_b\over6}\widehat g^{\,2}(\zeta-1)
 +O((\zeta-1)^2),
\label{eq:fermVelApp}\\
 {\dd\ln c\over\dd\ell}&=N_f\widehat g^{\,2}(\zeta-1)
 +O((\zeta-1)^2).
\label{eq:bosVelApp}
\end{align}
Their difference yields
\begin{equation}
 {\dd\zeta\over\dd\ell}=-\left(N_f+{N_b\over6}\right)
 \widehat g^{\,2}\zeta(\zeta-1)+O((\zeta-1)^2).
\label{eq:AzetaApp2}
\end{equation}
For four-component Dirac fermions at the chiral-XY fixed point, $N_f=N_b=2$ and $\widehat g^{\,2*}=\epsilon/12$, giving $\omega_\zeta=7\epsilon/36$ (see  Eq.~\eqref{eq:omega-zeta}).  This confirms stability of the equal-velocity point inside the isotropic continuum subspace~\cite{RoyHerbut2016}.

\section{Brazovskii fluctuation criterion}
\label{app:brazovskii}

A different situation arises when the quadratic kernel has an approximately
continuous ring of minima, $|\mathbf{k}|\simeq Q$, rather than an isolated
lattice-selected pair $\pm\mathbf Q$. The large phase space of nearly soft
modes can then preempt the continuous transition through the classical
Brazovskii mechanism~\cite{Brazovskii1975}. In the Hartree approximation, the
renormalized mass satisfies
\begin{equation}
 r=r_0+\mathcal B_Q\lambda_4\,r^{-1/2},
\label{eq:BrazSelf}
\end{equation}
Here $\mathcal B_Q$ is the phase-space factor associated with the ring of minima and $\lambda_4$ is the dimensionful quartic coefficient. Within the Hartree--Brazovskii treatment, the singular term prevents the renormalized mass from vanishing continuously. The instability occurs when the free energy of the state with a ripple becomes lower than that of the state with $\langle\Phi_Q\rangle=0$, leading to a fluctuation-driven first-order transition characterized parametrically by
\begin{equation}
 r_{\rm tr}\sim (\mathcal B_Q\lambda_4)^{2/3},
 \qquad
 \xi_{\max}\sim (\mathcal B_Q\lambda_4)^{-1/3}.
\end{equation}
This Hartree criterion establishes the breakdown of the continuous transition. In two dimensions, the long-distance structure of the modulated state can be further affected by orientational fluctuations and topological defects. When lattice anisotropy selects only the isolated pair $\pm\mathbf Q$, the ring degeneracy is removed and the relativistic finite-$Q$ critical theory of section~\ref{sec:finiteQ} applies.

At zero temperature, an approximately continuous ring of minima is less
singular than in the classical problem. Only the frequency $\omega$ and the
radial deviation $k_\perp$ from $|\mathbf{k}|=Q$ become soft, while motion
along the ring contributes a finite phase-space factor. The resulting
infrared contribution is
\begin{equation}
 \int\dd\omega\,\dd k_\perp\,
 \frac{1}{\omega^2+c^2k_\perp^2+r}
 \sim \ln\frac{\Lambda}{\sqrt r}.
\label{eq:quantumShellLog}
\end{equation}
Thus the zero-temperature ring problem is marginal by power counting and produces a logarithmic singularity rather than the $r^{-1/2}$ behavior of the classical Hartree problem. This logarithm alone does not determine whether the quantum transition is continuous or weakly first order; the flow of the angularly resolved quartic interactions requires a separate RG analysis. When lattice anisotropy selects only the isolated pair $\pm\mathbf Q$, the phase-space enhancement is absent and the relativistic finite-$Q$ theory of section~\ref{sec:finiteQ} applies.

\bibliographystyle{JHEP}
\bibliography{membranes_references}

@article{CastroNeto2009,
  author  = {Castro Neto, A. H. and Guinea, F. and Peres, N. M. R. and Novoselov, K. S. and Geim, A. K.},
  title   = {The electronic properties of graphene},
  journal = {Rev. Mod. Phys.},
  volume  = {81},
  number  = {1},
  pages   = {109--162},
  year    = {2009},
  doi     = {10.1103/RevModPhys.81.109}
}

@article{Roy-Juricic-PRB2014,
  title = {Strain-induced time-reversal odd superconductivity in graphene},
  author = {Roy, Bitan and Juri\ifmmode \check{c}\else \v{c}\fi{}i\ifmmode \acute{c}\else \'{c}\fi{}, Vladimir},
  journal = {Phys. Rev. B},
  volume = {90},
  issue = {4},
  pages = {041413(R)},
  numpages = {5},
  year = {2014},
  month = {Jul},
  publisher = {American Physical Society},
  doi = {10.1103/PhysRevB.90.041413},
  url = {https://link.aps.org/doi/10.1103/PhysRevB.90.041413}
}

@article{Isobe-Nagaosa-PRB2012,
  title = {Theory of a quantum critical phenomenon in a topological insulator: (3+1)-dimensional quantum electrodynamics in solids},
  author = {Isobe, Hiroki and Nagaosa, Naoto},
  journal = {Phys. Rev. B},
  volume = {86},
  issue = {16},
  pages = {165127},
  numpages = {5},
  year = {2012},
  month = {Oct},
  publisher = {American Physical Society},
  doi = {10.1103/PhysRevB.86.165127},
  url = {https://link.aps.org/doi/10.1103/PhysRevB.86.165127}
}

@article{Roy-Juricic-PRB-Kekule,
  title = {Fermionic multicriticality near Kekul\'e valence-bond ordering on a honeycomb lattice},
  author = {Roy, Bitan and Juri\ifmmode \check{c}\else \v{c}\fi{}i\ifmmode \acute{c}\else \'{c}\fi{}, Vladimir},
  journal = {Phys. Rev. B},
  volume = {99},
  issue = {24},
  pages = {241103(R)},
  numpages = {6},
  year = {2019},
  month = {Jun},
  publisher = {American Physical Society},
  doi = {10.1103/PhysRevB.99.241103},
  url = {https://link.aps.org/doi/10.1103/PhysRevB.99.241103}
}

@article{AssaadHerbut2013PRX,
  author        = {Assaad, Fakher F. and Herbut, Igor F.},
  title         = {Pinning the Order: The Nature of Quantum Criticality in the Hubbard Model on Honeycomb Lattice},
  journal       = {Phys. Rev. X},
  volume        = {3},
  number        = {3},
  pages         = {031010},
  year          = {2013},
  doi           = {10.1103/PhysRevX.3.031010},
  eprint        = {1304.6340},
  archivePrefix = {arXiv},
  primaryClass  = {cond-mat.str-el}
}

@article{OtsukaYunokiSorella2016PRX,
  author        = {Otsuka, Yuichi and Yunoki, Seiji and Sorella, Sandro},
  title         = {Universal Quantum Criticality in the Metal-Insulator Transition of Two-Dimensional Interacting Dirac Electrons},
  journal       = {Phys. Rev. X},
  volume        = {6},
  number        = {1},
  pages         = {011029},
  year          = {2016},
  doi           = {10.1103/PhysRevX.6.011029},
  eprint        = {1510.08593},
  archivePrefix = {arXiv},
  primaryClass  = {cond-mat.str-el}
}

@article{Ma2025RelativisticMott,
  author  = {Ma, Liguo and Chaturvedi, Raghav and Nguyen, Phuong X. and
             Watanabe, Kenji and Taniguchi, Takashi and Mak, Kin Fai and
             Shan, Jie},
  title   = {Relativistic {Mott} transition in twisted {WSe$_2$} tetralayers},
  journal = {Nature Materials},
  volume  = {24},
  number  = {12},
  pages   = {1935--1941},
  year    = {2025},
  doi     = {10.1038/s41563-025-02359-8}
}

@book{PeskinSchroeder1995,
  author    = {Peskin, Michael E. and Schroeder, Daniel V.},
  title     = {An Introduction to Quantum Field Theory},
  publisher = {Addison-Wesley},
  address   = {Reading, Massachusetts},
  year      = {1995},
  isbn      = {978-0-201-50397-5}
}

@article{AronovitzGolubovicLubensky1989,
  author  = {Aronovitz, Joseph A. and Golubovi{\'c}, Leonardo and
             Lubensky, T. C.},
  title   = {Fluctuations and Lower Critical Dimensions of Crystalline Membranes},
  journal = {J. Phys. France},
  volume  = {50},
  number  = {6},
  pages   = {609--631},
  year    = {1989},
  doi     = {10.1051/jphys:01989005006060900}
}

@article{RoyKennettYangJuricic2018,
  author        = {Roy, Bitan and Kennett, Malcolm P. and Yang, Kun and Juri{\v c}i{\'c}, Vladimir},
  title         = {From Birefringent Electrons to a Marginal or Non-Fermi Liquid of Relativistic Spin-$1/2$ Fermions: An Emergent Superuniversality},
  journal       = {Physical Review Letters},
  volume        = {121},
  number        = {15},
  pages         = {157602},
  year          = {2018},
  doi           = {10.1103/PhysRevLett.121.157602},
  eprint        = {1802.02134},
  archivePrefix = {arXiv},
  primaryClass  = {cond-mat.str-el}
}

@article{Wehling2014,
  author  = {Wehling, T. O. and Black-Schaffer, A. M. and Balatsky, A. V.},
  title   = {Dirac materials},
  journal = {Adv. Phys.},
  volume  = {63},
  number  = {1},
  pages   = {1--76},
  year    = {2014},
  doi     = {10.1080/00018732.2014.927109},
  eprint  = {1405.5774},
  archivePrefix = {arXiv},
  primaryClass  = {cond-mat.mtrl-sci}
}

@article{Armitage2018,
  author  = {Armitage, N. P. and Mele, E. J. and Vishwanath, A.},
  title   = {Weyl and Dirac semimetals in three-dimensional solids},
  journal = {Rev. Mod. Phys.},
  volume  = {90},
  number  = {1},
  pages   = {015001},
  year    = {2018},
  doi     = {10.1103/RevModPhys.90.015001},
  eprint  = {1705.01111},
  archivePrefix = {arXiv},
  primaryClass  = {cond-mat.str-el}
}

@article{Gonzalez1994,
  author  = {Gonz{\'a}lez, J. and Guinea, F. and Vozmediano, M. A. H.},
  title   = {Non-Fermi liquid behavior of electrons in the half-filled honeycomb lattice},
  journal = {Nucl. Phys. B},
  volume  = {424},
  number  = {3},
  pages   = {595--618},
  year    = {1994},
  doi     = {10.1016/0550-3213(94)90410-3}
}

@article{RoyHerbut2016,
  author  = {Roy, B. and Juri{\v{c}}i{\'c}, V. and Herbut, I. F.},
  title   = {Emergent Lorentz symmetry near fermionic quantum critical points in two and three dimensions},
  journal = {JHEP},
  volume  = {04},
  pages   = {018},
  year    = {2016},
  doi     = {10.1007/JHEP04(2016)018},
  eprint  = {1510.07650},
  archivePrefix = {arXiv},
  primaryClass  = {cond-mat.str-el}
}

@article{ReiserJuricic2024,
  author  = {Reiser, P. and Juri{\v{c}}i{\'c}, V.},
  title   = {Tilted Dirac superconductor at quantum criticality: restoration of Lorentz symmetry},
  journal = {JHEP},
  volume  = {02},
  pages   = {181},
  year    = {2024},
  doi     = {10.1007/JHEP02(2024)181}
}

@article{NelsonPeliti_JPhysique_1987,
  author  = {Nelson, D. R. and Peliti, L.},
  title   = {Fluctuations in membranes with crystalline and hexatic order},
  journal = {J. Phys. France},
  volume  = {48},
  number  = {7},
  pages   = {1085--1092},
  year    = {1987},
  doi     = {10.1051/jphys:019870048070108500}
}

@article{Aronovitz_PRL_1988,
  author  = {Aronovitz, J. A. and Lubensky, T. C.},
  title   = {Fluctuations of solid membranes},
  journal = {Phys. Rev. Lett.},
  volume  = {60},
  number  = {25},
  pages   = {2634--2637},
  year    = {1988},
  doi     = {10.1103/PhysRevLett.60.2634}
}

@article{LeDoussalRadzihovsky1992,
  author  = {Le Doussal, P. and Radzihovsky, L.},
  title   = {Self-consistent theory of polymerized membranes},
  journal = {Phys. Rev. Lett.},
  volume  = {69},
  number  = {8},
  pages   = {1209--1212},
  year    = {1992},
  doi     = {10.1103/PhysRevLett.69.1209}
}

@book{NelsonPiranWeinberg2004,
  editor    = {Nelson, D. R. and Piran, T. and Weinberg, S.},
  title     = {Statistical Mechanics of Membranes and Surfaces},
  edition   = {2},
  publisher = {World Scientific},
  address   = {Singapore},
  year      = {2004},
  doi       = {10.1142/5473}
}

@article{LeDoussalRadzihovsky2018,
  author  = {Le Doussal, P. and Radzihovsky, L.},
  title   = {Anomalous elasticity, fluctuations and disorder in elastic membranes},
  journal = {Ann. Phys.},
  volume  = {392},
  pages   = {340--410},
  year    = {2018},
  doi     = {10.1016/j.aop.2017.08.033},
  eprint  = {1708.05723},
  archivePrefix = {arXiv},
  primaryClass  = {cond-mat.stat-mech}
}

@article{Kownacki_PRE_2009,
  author  = {Kownacki, J.-P. and Mouhanna, D.},
  title   = {Crumpling transition and flat phase of polymerized phantom membranes},
  journal = {Phys. Rev. E},
  volume  = {79},
  number  = {4},
  pages   = {040101},
  year    = {2009},
  doi     = {10.1103/PhysRevE.79.040101}
}

@article{Coquand_PRE_2020,
  author  = {Coquand, O. and Mouhanna, D. and Teber, S.},
  title   = {Flat phase of polymerized membranes at two-loop order},
  journal = {Phys. Rev. E},
  volume  = {101},
  number  = {6},
  pages   = {062104},
  year    = {2020},
  doi     = {10.1103/PhysRevE.101.062104}
}

@article{MetayerTeber_JSTAT_2025,
  author  = {Metayer, S. and Teber, S.},
  title   = {Field-theory approach to flat polymerized membranes},
  journal = {J. Stat. Mech.},
  volume  = {2025},
  number  = {9},
  pages   = {092001},
  year    = {2025},
  doi     = {10.1088/1742-5468/add516},
  eprint  = {2412.18490},
  archivePrefix = {arXiv},
  primaryClass  = {cond-mat.stat-mech}
}

@article{MarianiOppen2008,
  author  = {Mariani, E. and von Oppen, F.},
  title   = {Flexural phonons in free-standing graphene},
  journal = {Phys. Rev. Lett.},
  volume  = {100},
  number  = {7},
  pages   = {076801},
  year    = {2008},
  doi     = {10.1103/PhysRevLett.100.076801},
  eprint  = {0707.4350},
  archivePrefix = {arXiv},
  primaryClass  = {cond-mat.mes-hall}
}

@article{Guinea_PRB_2014,
  author  = {Guinea, F. and Le Doussal, P. and Wiese, K. J.},
  title   = {Collective excitations in a large-$d$ model for graphene},
  journal = {Phys. Rev. B},
  volume  = {89},
  number  = {12},
  pages   = {125428},
  year    = {2014},
  doi     = {10.1103/PhysRevB.89.125428}
}

@article{MauriKatsnelson2022,
  author  = {Mauri, A. and Katsnelson, M. I.},
  title   = {Perturbative renormalization and thermodynamics of quantum crystalline membranes},
  journal = {Phys. Rev. B},
  volume  = {105},
  number  = {19},
  pages   = {195434},
  year    = {2022},
  doi     = {10.1103/PhysRevB.105.195434},
  eprint  = {2202.12842},
  archivePrefix = {arXiv},
  primaryClass  = {cond-mat.stat-mech}
}

@article{KatsnelsonFasolino2013,
  author  = {Katsnelson, M. I. and Fasolino, A.},
  title   = {Graphene as a prototype crystalline membrane},
  journal = {Acc. Chem. Res.},
  volume  = {46},
  number  = {1},
  pages   = {97--105},
  year    = {2013},
  doi     = {10.1021/ar300117m}
}

@article{KatsLebedev2014,
  author  = {Kats, E. I. and Lebedev, V. V.},
  title   = {Asymptotic freedom at zero temperature in free-standing crystalline membranes},
  journal = {Phys. Rev. B},
  volume  = {89},
  number  = {12},
  pages   = {125433},
  year    = {2014},
  doi     = {10.1103/PhysRevB.89.125433}
}

@article{Amorim2014,
  author  = {Amorim, B. and Rold{\'a}n, R. and Cappelluti, E. and Fasolino, A. and Guinea, F. and Katsnelson, M. I.},
  title   = {Thermodynamics of quantum crystalline membranes},
  journal = {Phys. Rev. B},
  volume  = {89},
  number  = {22},
  pages   = {224307},
  year    = {2014},
  doi     = {10.1103/PhysRevB.89.224307},
  eprint  = {1403.2635},
  archivePrefix = {arXiv},
  primaryClass  = {cond-mat.mtrl-sci}
}

@article{GonzalezPerfetto2009,
  author  = {Gonz{\'a}lez, J. and Perfetto, E.},
  title   = {Many-body effects on out-of-plane phonons in graphene},
  journal = {New J. Phys.},
  volume  = {11},
  pages   = {095015},
  year    = {2009},
  doi     = {10.1088/1367-2630/11/9/095015},
  eprint  = {0906.4969},
  archivePrefix = {arXiv},
  primaryClass  = {cond-mat.mes-hall}
}

@article{SanJoseGonzalezGuinea2011,
  author  = {San-Jose, P. and Gonz{\'a}lez, J. and Guinea, F.},
  title   = {Electron-induced rippling in graphene},
  journal = {Phys. Rev. Lett.},
  volume  = {106},
  number  = {4},
  pages   = {045502},
  year    = {2011},
  doi     = {10.1103/PhysRevLett.106.045502},
  eprint  = {1009.4588},
  archivePrefix = {arXiv},
  primaryClass  = {cond-mat.mes-hall}
}

@article{GuitterDavidLeiblerPeliti1989,
  author  = {Guitter, E. and David, F. and Leibler, S. and Peliti, L.},
  title   = {Thermodynamical behavior of polymerized membranes},
  journal = {J. Phys. France},
  volume  = {50},
  number  = {14},
  pages   = {1787--1819},
  year    = {1989},
  doi     = {10.1051/jphys:0198900500140178700}
}

@article{RoyJuricicHerbut2013,
  author        = {Roy, Bitan and Juri\v{c}i\'c, Vladimir and Herbut, Igor F.},
  title         = {Quantum superconducting criticality in graphene and topological insulators},
  journal       = {Phys. Rev. B},
  volume        = {87},
  number        = {4},
  pages         = {041401},
  year          = {2013},
  doi           = {10.1103/PhysRevB.87.041401},
  eprint        = {1210.3576},
  archivePrefix = {arXiv},
  primaryClass  = {cond-mat.str-el},
  note          = {Erratum: Phys. Rev. B 94, 119901 (2016)}
}

@article{GraceyMaierMarquardSchroder2025,
  author        = {Gracey, J. A. and Maier, A. and Marquard, P. and Schr{\"o}der, Y.},
  title         = {Anomalous dimensions and critical exponents for the
                   {Gross--Neveu--Yukawa} model at five loops},
  journal       = {Phys. Rev. D},
  volume        = {112},
  pages         = {085029},
  year          = {2025},
  doi           = {10.1103/lmpx-n3mj},
  eprint        = {2507.22594},
  archivePrefix = {arXiv},
  primaryClass  = {hep-th}
}

@article{Gracey2021ChiralXY,
  author        = {Gracey, J. A.},
  title         = {Critical exponent $\eta$ at $O(1/N^3)$ in the chiral
                   {XY} model using the large-$N$ conformal bootstrap},
  journal       = {Phys. Rev. D},
  volume        = {103},
  pages         = {065018},
  year          = {2021},
  doi           = {10.1103/PhysRevD.103.065018},
  eprint        = {2101.03385},
  archivePrefix = {arXiv},
  primaryClass  = {hep-th}
}

@article{FeiGiombiKlebanovTarnopolsky2016,
  author        = {Fei, Lin and Giombi, Simone and Klebanov, Igor R. and Tarnopolsky, Grigory},
  title         = {Yukawa conformal field theories and emergent supersymmetry},
  journal       = {Prog. Theor. Exp. Phys.},
  volume        = {2016},
  number        = {12},
  pages         = {12C105},
  year          = {2016},
  doi           = {10.1093/ptep/ptw120},
  eprint        = {1607.05316},
  archivePrefix = {arXiv},
  primaryClass  = {hep-th}
}

@article{Gonzalez2014,
  author  = {Gonz{\'a}lez, J.},
  title   = {Rippling transition from electron-induced condensation of curvature field in graphene},
  journal = {Phys. Rev. B},
  volume  = {90},
  number  = {16},
  pages   = {165402},
  year    = {2014},
  doi     = {10.1103/PhysRevB.90.165402},
  eprint  = {1407.7545},
  archivePrefix = {arXiv},
  primaryClass  = {cond-mat.mes-hall}
}

@article{Brazovskii1975,
  author  = {Brazovskii, S. A.},
  title   = {Phase transition of an isotropic system to a nonuniform state},
  journal = {Sov. Phys. JETP},
  volume  = {41},
  number  = {1},
  pages   = {85--89},
  year    = {1975}
}

@book{ZinnJustin2002,
  author    = {Zinn-Justin, Jean},
  title     = {Quantum Field Theory and Critical Phenomena},
  edition   = {4},
  publisher = {Oxford University Press},
  address   = {Oxford},
  year      = {2002}
}

@article{HouChamonMudry2007,
  author  = {Hou, C.-Y. and Chamon, C. and Mudry, C.},
  title   = {Electron fractionalization in two-dimensional graphenelike structures},
  journal = {Phys. Rev. Lett.},
  volume  = {98},
  number  = {18},
  pages   = {186809},
  year    = {2007},
  doi     = {10.1103/PhysRevLett.98.186809},
  eprint  = {cond-mat/0609740},
  archivePrefix = {arXiv}
}

@article{HerbutJuricicVafek2009,
  author  = {Herbut, I. F. and Juri{\v{c}}i{\'c}, V. and Vafek, O.},
  title   = {Relativistic Mott criticality in graphene},
  journal = {Phys. Rev. B},
  volume  = {80},
  number  = {7},
  pages   = {075432},
  year    = {2009},
  doi     = {10.1103/PhysRevB.80.075432},
  eprint  = {0904.1019},
  archivePrefix = {arXiv},
  primaryClass  = {cond-mat.str-el}
}

@article{LiJiangJianYao2017,
  author  = {Li, Z.-X. and Jiang, Y.-F. and Jian, S.-K. and Yao, H.},
  title   = {Fermion-induced quantum critical points},
  journal = {Nat. Commun.},
  volume  = {8},
  pages   = {314},
  year    = {2017},
  doi     = {10.1038/s41467-017-00167-6},
  eprint  = {1512.07908},
  archivePrefix = {arXiv},
  primaryClass  = {cond-mat.str-el}
}

@article{BowickTravesset2001,
  author        = {Bowick, Mark J. and Travesset, Alex},
  title         = {The statistical mechanics of membranes},
  journal       = {Phys. Rep.},
  volume        = {344},
  number        = {4--6},
  pages         = {255--308},
  year          = {2001},
  doi           = {10.1016/S0370-1573(00)00128-9},
  eprint        = {cond-mat/0002038},
  archivePrefix = {arXiv}
}

@article{Gazit2009,
  author  = {Gazit, Doron},
  title   = {Structure of physical crystalline membranes within the self-consistent screening approximation},
  journal = {Phys. Rev. E},
  volume  = {80},
  number  = {4},
  pages   = {041117},
  year    = {2009},
  doi     = {10.1103/PhysRevE.80.041117}
}

@article{RoldanFasolinoZakharchenkoKatsnelson2011,
  author  = {Rold{\'a}n, Rafael and Fasolino, Annalisa and Zakharchenko, Kostyantyn V. and Katsnelson, Mikhail I.},
  title   = {Suppression of anharmonicities in crystalline membranes by external strain},
  journal = {Phys. Rev. B},
  volume  = {83},
  number  = {17},
  pages   = {174104},
  year    = {2011},
  doi     = {10.1103/PhysRevB.83.174104}
}

@article{MauriKatsnelson2020,
  author  = {Mauri, Andrea and Katsnelson, Mikhail I.},
  title   = {Scaling behavior of crystalline membranes: An $\epsilon$-expansion approach},
  journal = {Nucl. Phys. B},
  volume  = {956},
  pages   = {115040},
  year    = {2020},
  doi     = {10.1016/j.nuclphysb.2020.115040}
}

@article{BurmistrovStress2018,
  author        = {Burmistrov, I. S. and Gornyi, I. V. and Kachorovskii, V. Yu. and Katsnelson, M. I. and Los, J. H. and Mirlin, A. D.},
  title         = {Stress-controlled Poisson ratio of a crystalline membrane: Application to graphene},
  journal       = {Phys. Rev. B},
  volume        = {97},
  number        = {12},
  pages         = {125402},
  year          = {2018},
  doi           = {10.1103/PhysRevB.97.125402},
  eprint        = {1801.05476},
  archivePrefix = {arXiv},
  primaryClass  = {cond-mat.mes-hall}
}

@article{BurmistrovDifferential2018,
  author        = {Burmistrov, I. S. and Kachorovskii, V. Yu. and Gornyi, I. V. and Mirlin, A. D.},
  title         = {Differential Poisson's ratio of a crystalline two-dimensional membrane},
  journal       = {Ann. Phys.},
  volume        = {396},
  pages         = {119--136},
  year          = {2018},
  doi           = {10.1016/j.aop.2018.07.009},
  eprint        = {1801.05053},
  archivePrefix = {arXiv},
  primaryClass  = {cond-mat.mes-hall}
}

@article{VozmedianoKatsnelsonGuinea2010,
  author        = {Vozmediano, M. A. H. and Katsnelson, M. I. and Guinea, F.},
  title         = {Gauge fields in graphene},
  journal       = {Phys. Rep.},
  volume        = {496},
  number        = {4--5},
  pages         = {109--148},
  year          = {2010},
  doi           = {10.1016/j.physrep.2010.07.003},
  eprint        = {1003.5179},
  archivePrefix = {arXiv},
  primaryClass  = {cond-mat.mes-hall}
}

@article{Manes2007,
  author  = {Ma{\~n}es, J. L.},
  title   = {Symmetry-based approach to electron-phonon interactions in graphene},
  journal = {Phys. Rev. B},
  volume  = {76},
  number  = {4},
  pages   = {045430},
  year    = {2007},
  doi     = {10.1103/PhysRevB.76.045430}
}

@article{GuineaHorovitzLeDoussal2008,
  author  = {Guinea, F. and Horovitz, Baruch and Le Doussal, P.},
  title   = {Gauge field induced by ripples in graphene},
  journal = {Phys. Rev. B},
  volume  = {77},
  number  = {20},
  pages   = {205421},
  year    = {2008},
  doi     = {10.1103/PhysRevB.77.205421}
}

@article{WunschStauberSolsGuinea2006,
  author        = {Wunsch, B. and Stauber, T. and Sols, F. and Guinea, F.},
  title         = {Dynamical polarization of graphene at finite doping},
  journal       = {New J. Phys.},
  volume        = {8},
  pages         = {318},
  year          = {2006},
  doi           = {10.1088/1367-2630/8/12/318},
  eprint        = {cond-mat/0610630},
  archivePrefix = {arXiv}
}

@article{HwangDasSarma2007,
  author  = {Hwang, E. H. and Das Sarma, S.},
  title   = {Dielectric function, screening, and plasmons in two-dimensional graphene},
  journal = {Phys. Rev. B},
  volume  = {75},
  number  = {20},
  pages   = {205418},
  year    = {2007},
  doi     = {10.1103/PhysRevB.75.205418}
}

@article{GuineaKatsnelsonGeim2010,
  author  = {Guinea, F. and Katsnelson, M. I. and Geim, A. K.},
  title   = {Energy gaps and a zero-field quantum Hall effect in graphene by strain engineering},
  journal = {Nat. Phys.},
  volume  = {6},
  number  = {1},
  pages   = {30--33},
  year    = {2010},
  doi     = {10.1038/nphys1420}
}

@article{McMillan1976,
  author  = {McMillan, W. L.},
  title   = {Theory of discommensurations and the commensurate-incommensurate charge-density-wave phase transition},
  journal = {Phys. Rev. B},
  volume  = {14},
  number  = {4},
  pages   = {1496--1502},
  year    = {1976},
  doi     = {10.1103/PhysRevB.14.1496}
}

@article{Bak1982,
  author  = {Bak, P.},
  title   = {Commensurate phases, incommensurate phases and the devil's staircase},
  journal = {Rep. Prog. Phys.},
  volume  = {45},
  number  = {6},
  pages   = {587--629},
  year    = {1982},
  doi     = {10.1088/0034-4885/45/6/001}
}

@article{SwiftHohenberg1977,
  author  = {Swift, J. and Hohenberg, P. C.},
  title   = {Hydrodynamic fluctuations at the convective instability},
  journal = {Phys. Rev. A},
  volume  = {15},
  number  = {1},
  pages   = {319--328},
  year    = {1977},
  doi     = {10.1103/PhysRevA.15.319}
}

@article{HerbutJuricicRoy2009,
  author        = {Herbut, Igor F. and Juri{\v{c}}i{\'c}, Vladimir and Roy, Bitan},
  title         = {Theory of interacting electrons on the honeycomb lattice},
  journal       = {Phys. Rev. B},
  volume        = {79},
  number        = {8},
  pages         = {085116},
  year          = {2009},
  doi           = {10.1103/PhysRevB.79.085116},
  eprint        = {0811.0610},
  archivePrefix = {arXiv},
  primaryClass  = {cond-mat.str-el}
}

@article{ZerfLinMaciejko2016,
  author        = {Zerf, Nikolai and Lin, Chien-Hung and Maciejko, Joseph},
  title         = {Superconducting quantum criticality of topological surface states at three loops},
  journal       = {Phys. Rev. B},
  volume        = {94},
  number        = {20},
  pages         = {205106},
  year          = {2016},
  doi           = {10.1103/PhysRevB.94.205106},
  eprint        = {1605.09423},
  archivePrefix = {arXiv},
  primaryClass  = {cond-mat.str-el}
}

@article{OtsukaSekiSorellaYunoki2018,
  author        = {Otsuka, Yuichi and Seki, Kazuhiro and Sorella, Sandro and Yunoki, Seiji},
  title         = {Quantum criticality in the metal-superconductor transition of interacting {Dirac} fermions on a triangular lattice},
  journal       = {Phys. Rev. B},
  volume        = {98},
  number        = {3},
  pages         = {035126},
  year          = {2018},
  doi           = {10.1103/PhysRevB.98.035126},
  eprint        = {1803.02001},
  archivePrefix = {arXiv},
  primaryClass  = {cond-mat.str-el}
}

@article{LiLiYao2020,
  author        = {Li, Bo-Hai and Li, Zi-Xiang and Yao, Hong},
  title         = {Fermion-induced quantum critical point in {Dirac} semimetals: A sign-problem-free quantum {Monte Carlo} study},
  journal       = {Phys. Rev. B},
  volume        = {101},
  number        = {8},
  pages         = {085105},
  year          = {2020},
  doi           = {10.1103/PhysRevB.101.085105},
  eprint        = {1910.14287},
  archivePrefix = {arXiv},
  primaryClass  = {cond-mat.str-el}
}

@article{LiuWangSunMeng2020,
  author        = {Liu, Yuzhi and Wang, Wei and Sun, Kai and Meng, Zi Yang},
  title         = {Designer {Monte Carlo} simulation for the {Gross--Neveu--Yukawa} transition},
  journal       = {Phys. Rev. B},
  volume        = {101},
  number        = {6},
  pages         = {064308},
  year          = {2020},
  doi           = {10.1103/PhysRevB.101.064308},
  eprint        = {1910.07430},
  archivePrefix = {arXiv},
  primaryClass  = {cond-mat.stat-mech}
}

@article{Campostrini2006,
  author        = {Campostrini, Massimo and Hasenbusch, Martin and Pelissetto, Andrea and Vicari, Ettore},
  title         = {The critical exponents of the superfluid transition in ${}^{4}$He},
  journal       = {Phys. Rev. B},
  volume        = {74},
  number        = {14},
  pages         = {144506},
  year          = {2006},
  doi           = {10.1103/PhysRevB.74.144506},
  eprint        = {cond-mat/0605083},
  archivePrefix = {arXiv},
  primaryClass  = {cond-mat.stat-mech}
}
\end{document}